\documentclass[]{interact}
\usepackage{epstopdf}
\usepackage[caption=false]{subfig}

\usepackage[longnamesfirst,sort]{natbib}
\usepackage{makecell}
\usepackage[hyphens]{url}
\usepackage[hidelinks]{hyperref}
\hypersetup{breaklinks=true}
\usepackage{tablefootnote}
\bibpunct[, ]{(}{)}{;}{a}{,}{,}
\renewcommand\bibfont{\fontsize{10}{12}\selectfont}

\theoremstyle{plain}

\theoremstyle{definition}

\theoremstyle{remark}

\begin{document}

\articletype{RESEARCH ARTICLE}
\title{The Value-Sensitive Conversational Agent Co-Design Framework}

\author{
\name{Malak Sadek\textsuperscript{a}\thanks{CONTACT Malak Sadek. Email: m.sadek21@imperial.ac.uk} and Rafael A. Calvo\textsuperscript{a} and Celine Mougenot\textsuperscript{a}}
\affil{\textsuperscript{a} Dyson School of Design Engineering, Imperial College London, United Kingdom}
}

\maketitle
\begin{abstract}
Conversational agents (CAs) are gaining traction in both industry and academia, especially with the advent of generative
AI and large language models. As these agents are used more broadly by members of the general public and take on a number of critical use-cases and social roles, it becomes important to consider the values embedded in these systems. This consideration includes answering questions such as `whose values get embedded in these agents?' and `how do those values manifest in the agents being designed?' Accordingly, the aim of this paper is to present the Value-Sensitive Conversational Agent (VSCA) Framework for enabling the collaborative design (co-design) of value-sensitive CAs with relevant stakeholders. Firstly, requirements for co-designing value-sensitive CAs which were identified in previous works are summarised here. Secondly, the practical framework is presented and discussed, including its operationalisation into a design toolkit. The framework facilitates the co-design of three artefacts that elicit stakeholder values and have a technical utility to CA teams to guide CA implementation, enabling the creation of value embodied CA prototypes. Finally, an evaluation protocol for the framework is proposed where the effects of the framework and toolkit are explored in a design workshop setting to evaluate both the process followed and outcomes produced. 
\end{abstract}


\section{Introduction}
\label{sec:introduction}

Efforts to further technological advances in the space of Artificial Intelligence (AI) systems show no sign of slowing down, despite numerous requests and warnings from experts \citep{ai-warning}. Conversational agents (CAs) are one form of technology that commonly leverage AI to deliver conversational experiences through voice or text \citep{delloiteCA}. These types of systems have seen an exponential increase in interest, research, and use \citep{wahde2022}, even more so recently due to the proliferation and commercialisation of Generative AI and Large Language Models such as ChatGPT, earning them the title of ``Trojan horses for AI" ~[p. 68]\citep{russell2019}. 

\bigskip
\noindent Instead of halting development and progress altogether, another strong sentiment within the research community is to instead focus on ensuring these systems are human-centred \citep{crawford2016, lewandowski2021}. Such a focus can help offset and mitigate a number of harms caused by issues ranging from a reliance on biased data \citep{klein2020}, to a lack of general lack of explainability and transparency in the majority of systems developed \citep{digitalcatapult2020}. Recommendations for more human-centred AI have also extended to generative AI models and the systems that utilise them \citep{chen2023}. CAs have the potential to foster more inclusivity and accessibility over traditional graphical user interfaces by accommodating for a number of disabilities and impairments \citep{spatz2018} and providing crticial services and information in an accessible and affordable form \citep{folstad2018}. They can also enable ``heads-up computing” that moves beyond the device-centred paradigm to more human-centred interactions which focus on humans’ capabilities and resources, as opposed to forcing them to fit into technologies’ interaction modes which can be distracting, intrusive, and uncomfortable \citep{zhao2023}. 

\bigskip
\noindent Nevertheless, the biases present in other AI-based systems exist in CAs just the same. AINow Institute has already identified voice technologies being biased against women by having lower voice recognition accuracy for women’s voices, providing lacking health service information specific to women’s health, and stereo-typically grouping between women and occupations such as `secretary’ \citep{ainow2019}. Text-based chatbots have also shown instances where they responded more efficiently to words associated with males such as `wallet' over those associated to females such as `purse' \citep{}. Commercial users have also reported having to sound ``intentionally white” for voice-enabled CAs to understand them correctly \citep{byrd2020}. 

\bigskip
\noindent This increase in use and influence comes hand in hand with an increased responsibility to understand and manage the values embedded within these agents. In order to do so, practical mechanisms that allow for the involvement of a diverse range of stakeholders in order to collect and understand their values are needed \citep{dantec2009}. Mechanisms for effectively embedding those values into the technology being developed are also needed \citep{manders2011, poel2020}. Nevertheless, there is currently limited stakeholder participation and grounding in value-sensitivity across CAs \citep{byrd2020, folstad2021, showkat2022}.

\bigskip
\noindent Echoing the sentiments for needing to focus on human-centered AI, our research works towards re-imagining the design process for CAs in a way that (i) leverages collaborative design (co-design) with diverse stakeholders \citep{strappers2008} and (ii) focuses on creating value-sensitive outcomes \citep{friedman2006}. These two distinct design approaches are thus brought together where co-design is adapted and used as a means to enable value-sensitive outcomes. On the one hand, the goal of collaboratively designing CA is rooted in (i) well-cited benefits of co-designing technology in general \citep{strappers2008, Hayes2020}, and CAs specifically \citep{} (ii) calls and recommendations from experts regarding co-designing CAs \citep{chen2020, lewandowski2021}, and (iii) the necessity of involving relevant stakeholders in order to elicit their values and have them verify that those values have been successfully embedded in the CA. On the other hand, the creation of value-sensitive CAs stems from (i) the need to take control of value-based decision making in AI systems in terms of whose values get embedded and how (often referred to as the ``value alignment problem" \citep{christian2020}), (ii) the importance of value considerations in both AI-based systems \citep{umbrello2021} and CAs specifically \citep{wambsganss2021}, and (iii) the lack of emphasis on values in efforts to co-design CAs so far \citep{}. These two goals go hand-in-hand and form the basis of the research conducted.

\begin{figure}
    \centering
    \includegraphics[width=0.55\textwidth]{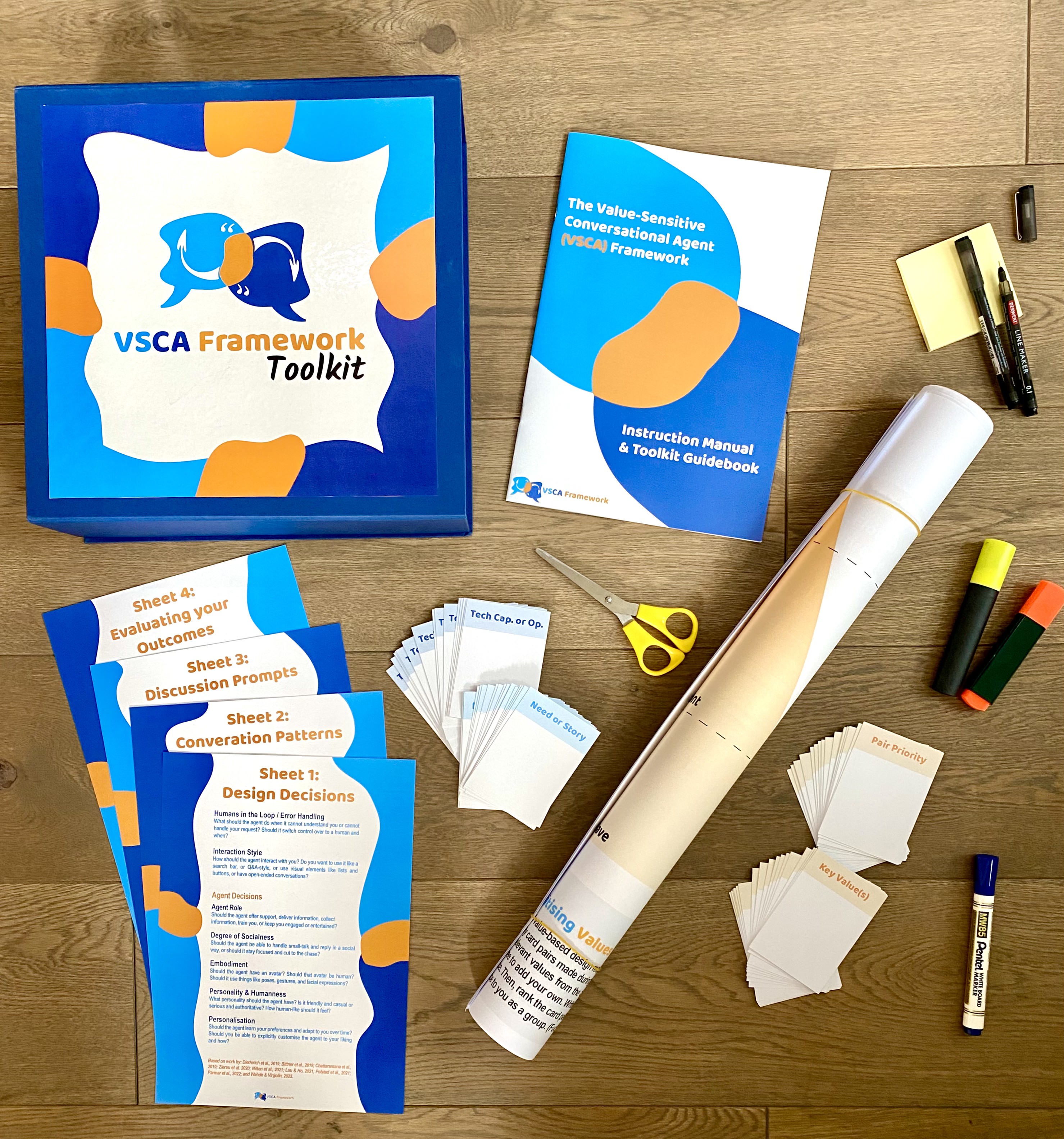}
    \caption{The VSCA Toolkit which operationalises the VSCA Framework. It consists of a guidebook, a set of canvases, information sheets, and blank decks of cards.}
    \label{fig:toolkit-intro}
\end{figure}

\bigskip
\noindent In this paper we present the Value-Sensitive Conversational Agent Co-Design Framework (VSCA). Figure \ref{fig:toolkit-intro} shows the toolkit created to operationalise the VSCA Framework. The framework is the result of a research process aimed at answering the question: ``how can we enable the co-design of value-sensitive conversational agents?".

\noindent Overall, the main objectives of the research were as follows:
\begin{enumerate}
    \item Identify requirements for co-designing value-sensitive CAs.
    \item Design and implement a practical framework to facilitate co-designing value-sensitive CAs.
    \item Explore the impact of the framework on resulting CA prototypes in a design workshop setting with teams building CAs.
\end{enumerate}

\begin{figure}
    \centering
    \includegraphics[width=\textwidth]{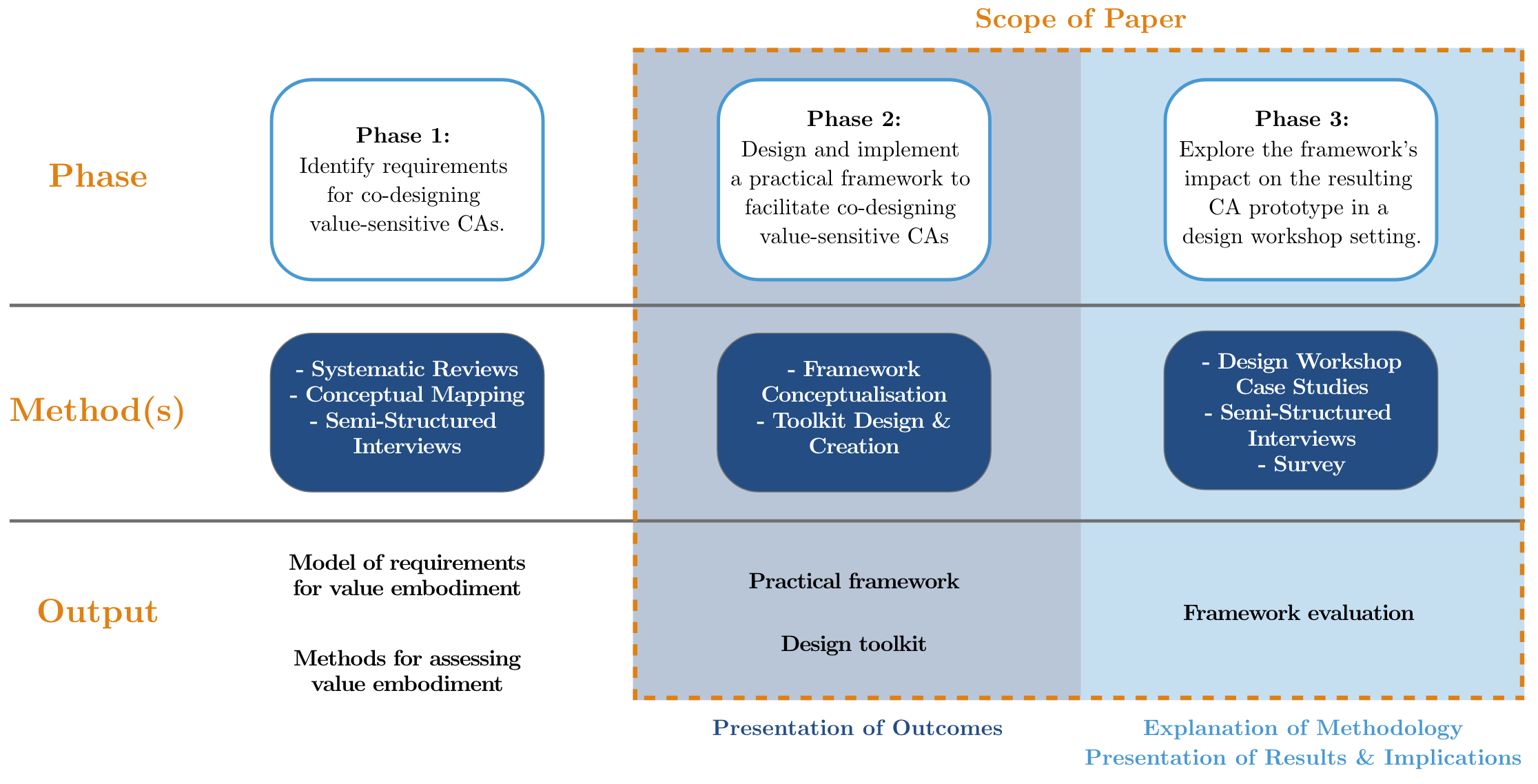}
    \caption{A diagram of the three project phases which covered the three objectives mentioned earlier highlighting the scope of this paper.}
    \label{fig:project-structure}
\end{figure}

\bigskip
\noindent These objectives shaped the three phases of this research as shown in Figure \ref{fig:project-structure}. Following these phases, after presenting the framework we will report on a case study where a team building a CA use the framework in a co-design workshop with their stakeholders. Finally, we propose two studies to explore the impact of the framework. The first study is a series of semi-structured interviews with workshop participants to understand their experiences with, and perceptions of, the framework. The second study is a survey administered to the target users of the CA designed during the workshop, in order to assess the value-sensitivity of the outcomes produced using the framework.

\section{Related Work}
\label{sec:background}

This section explores some of the discourse and literature that has led to the creation of the VSCA framework. First, the need for a practical framework over other forms of interventions is explored. Secondly, the use of boundary objects during AI systems design is documented. This is because the framework revolves around the creation of three boundary objects that embody stakeholders' values and can drive CA implementation forward. Thirdly, achieving and measuring value-sensitivity is discussed. This discussion helps contextualise the framework's rooting in value-sensitivity and the methodology for its evaluation. Finally, the requirements for co-designing value-sensitive CAs are presented as determined by previous work in order to set the stage for the framework's objectives and structure.

\subsection{Socio-Technical Interventions for AI Systems}

Within the realm of AI systems, researchers have recently called for a move away from abstract guidelines and principles that have a technical focus, and more towards concrete practices and toolkits that have socio-technical focuses \citep{harbers2022}. Such a recommendation is based on a number of prevalent criticisms regarding high-level guidelines and principles, including their abstract nature which leaves their interpretation up to practitioners \citep{whittlestone2019b, whittlestone2019c}, the difficulty of measuring their impact \citep{hagendorff2020} and the possibility that they may have no impact at all \citep{mcnamara2018}, and their tendency to promote passive compliance over active consideration of socio-technical aspects \citep{digitalcatapult2020, wachter2019}. Another problem is the over-saturation of available guidelines and recommendations, with studies finding over 80 documents offering guidelines for AI-based systems \citep{jobin2019} and over 300 guidelines for voice user interfaces \citep{murad2023}. This over-saturation makes it difficult to understand the differences between sources and choose an appropriate one. Finally, guidelines and recommendations were also found to be ineffective at handling several value-related aspects \citep{palmer2023}. Having a solely technical focus also leads to a number of limitations in AI systems, including a ``contextual awareness" \citep{cherubini2017} and a lack of support from stakeholders \citep{abaza2021}.

\bigskip
\noindent On the other hand, existing studies that focus on specific tools or design activities have suffered from a lack of standardisation \citep{klein2020} and limited coverage of the design life-cycle \citep{yildirim2023}. Moreover, in the space of CA design, experts have discussed the ``shortage of integrative perspectives on CA development and design" \citep{elshan2022}. Co-design is a valuable practice for democratising AI design \citep{koster2022} and foregrounding stakeholders' experiences and needs \citep{strappers2008} by treating them as ``experiential experts" \citep{eubanks2018}. While several studies focusing on co-designing different aspects of CAs exist, they lack a focus on stakeholder values and largely vary in terms of the relevance of stakeholders involved and the creation of prototypes \citep{}, which is needed for assessing whether values have been effectively embedded \citep{poel2020}.

\begin{figure}
    \centering
    \includegraphics[width=\textwidth]{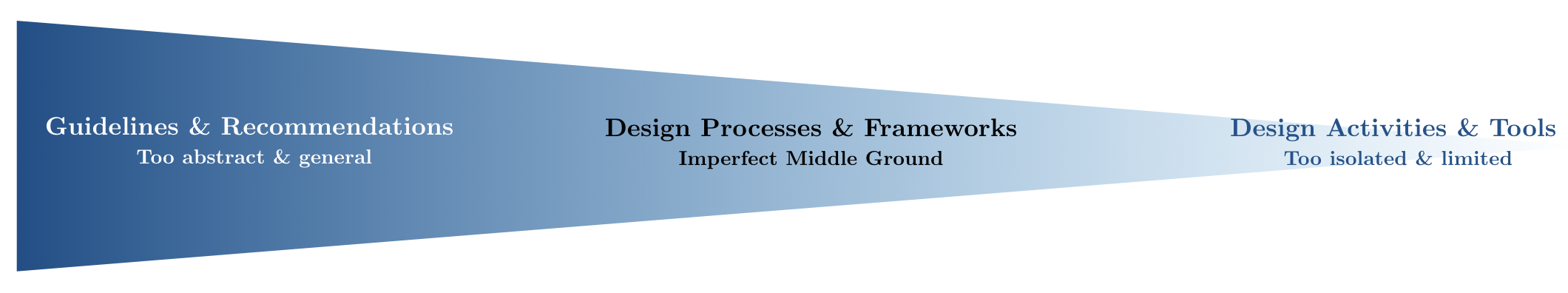}
    \caption{A spectrum depicting guidelines and recommendations as one extreme and design activities and tools as the other extreme, with design processes and frameworks as an imperfect middle ground.}
    \label{fig:spectrum}
\end{figure}

\bigskip
\noindent The selection of a practical framework as an intervention was rooted in the framing of design processes and frameworks as a middle ground, as shown in Figure \ref{fig:spectrum}. Nevertheless, recent work has shown that socio-technical design processes for AI systems vary in their capability to to support value-sensitive outcomes \citep{}. Other studies have also found that CA practitioners do to not use research-based design processes widely and that each practitioner largely follows their own process using ``trial and error" \citep{sadek2023}. Frameworks have the power to act as a middle ground between both extremes, benefiting from a broader perspective and more coverage than isolated activities or tools, and the ability to structure those activities and tools in a standardised and concrete way, avoiding the pitfalls of overly-general guidelines and principles. A framework can also be incorporated into practitioners' existing processes and workflows to achieve the framework's objective without redesigning their process from scratch or disrupting their workflow, thus facing less resistance to adoption from practitioners \citep{morley2021}. There have been also been calls from experts and practitioners for socio-technical frameworks for AI-based systems \citep{morley2021} and CAs \citep{lewandowski2021}.

\subsection{The Use of Boundary Objects in Designing AI Systems}

\noindent Creating boundary objects, also referred to as intermediate objects \citep{ocnarescu2011} or knowledge integrators \citep{lewandowski2021}, is a widely used technique to help reach a shared understanding or establish a shared model of reality between participants with differing perspectives or backgrounds \citep{pennington2010, cabitza2012, arrighi2019}. In interdisciplinary settings especially, a boundary object can act as (i) a tangible affordance of the knowledge produced, shared, and consumed throughout the activity \citep{pennington2010, cabitza2012, arrighi2019}, (ii) a form of documentation and basis for future-decision making, and (iii) a medium for discussions and conversations \citep{vines2013}. Design workshops often revolve around the creation and/or use of such boundary objects \citep{vines2013, steen2013, jonas2018}. Boundary objects can be closed objects that present or prescript information, such as models or design plans. Conversely, they can be open objects that act as inspiration or encourage interpretation and discussions such as sketches, decks of cards, and tangible objects \citep{redaelli2018, paavola2019}. Tangible objects and toolkits (such as canvases, decks of cards, figurines, sticky notes, etc.) have been found to be suitable for quick changes and fast evaluation \citep{boujut2003}.  

\bigskip
\noindent Another strength of boundary objects is their ability to act as a form of documentation. It is frequently cited that a significant limitation of design processes is the lack of documentation of, and reflection on, the knowledge produced and the design decisions taken \citep{hockey2007, sleeswijk2018, koesten2019}. This lack of documentation can lead to data going missing and confusion over team roles, expected outcomes, and the processes taking place - especially in interdisciplinary settings \citep{sleeswijk2018, koesten2019}. 

\bigskip
\noindent Looking at AI systems, where practitioners also complain of a lack of documentation, creating boundary objects can be especially important for transparency and accountability \citep{custis2021, fair2021}. Research has also found that AI practitioners use elements such as PowerPoint presentations as boundary objects, but find that they are time-consuming and do not provide sufficient documentation \citep{piorkowski2021}. Moreover, they are only made \textit{by} practitioners \textit{for} stakeholders (ibid). Conversely, \cite{yildirim2023} specifically recommend creating boundary objects that can be used by different stakeholders and team members. The use of boundary objects has also been recommended for in CA design \citep{lewandowski2021} and explored \citep{axelsson2021}, including whether CAs themselves can act as boundary objects \citep{kot2020}. User experience (UX) designers have also mentioned suffering from a lack of design materials and boundary objects for conversational UX design \citep{heo2023}. 

\subsection{Achieving and Measuring Value-Sensitivity}

Exploring value-related aspects of technology is far from new. Value-Sensitive Design (VSD) has been encouraging consideration of stakeholders' values on conceptual, empirical and technical levels when building technology for several years \citep{friedman2006}. More recently, researchers have worked on applying value-sensitive design to AI systems \citep{umbrello2021} and on considering other value-related aspects, such as whose values should be embedded in AI systems \citep{christian2020}, and what embedding values in AI systems could look like \citep{poel2020}. These works are rooted in the premise that a focus on values can bring numerous benefits to the development of AI systems including promoting reflexivity in practitioners \citep{umbrello2021}, increasing credibility and accountability \citep{stray2020, park2022}, and encouraging a society-in-the-loop approach with increased community participation \citep{rahwan2018}. 

\bigskip
\noindent As of yet, there has not been any work done on specifically applying VSD or other value-related frameworks to CAs, despite the existence of work regarding developing design guidelines for value-sensitive CAs \citep{wambsganss2021} or creating a value-sensitive CA with a specific group of stakeholders \citep{showkat2022}. 

\bigskip
\noindent Conceptually, working towards value-sensitive technology has been well-documented, including specific steps that need to take place throughout the AI system design process \citep{umbrello2021}. However, evaluating or measuring the value-sensitivity of a resulting prototype or system is challenging (as acknowledged by the creators of VSD \citep{friedman2021}). Given the abstract and subjective nature of values, it can be difficult to assess whether a technology enables a given value. One related approach used in the context of AI ethics is for AI creators or AI experts to self-assess how the system built fosters different AI ethics values \citep{yurrita2022, hallensleben2020}. The only existing work that explicitly assesses value-sensitivity is that of \cite{strikwerda2022}. They verify the value-sensitivity of their outcomes through an ethics matrix \citep{mepham2006}. This matrix is filled out by the researchers based on identified values emerging from discussions between target users. Such methods are a step in the right direction, but can fall prey to researcher subjectivity and a lack of empirical rigor. 

\subsection{Requirements for Co-Designing Value-Sensitive Conversational Agents}

\subsubsection{Reaching Value Embodiment}
\begin{figure}
    \centering
    \includegraphics[width=0.5\textwidth]{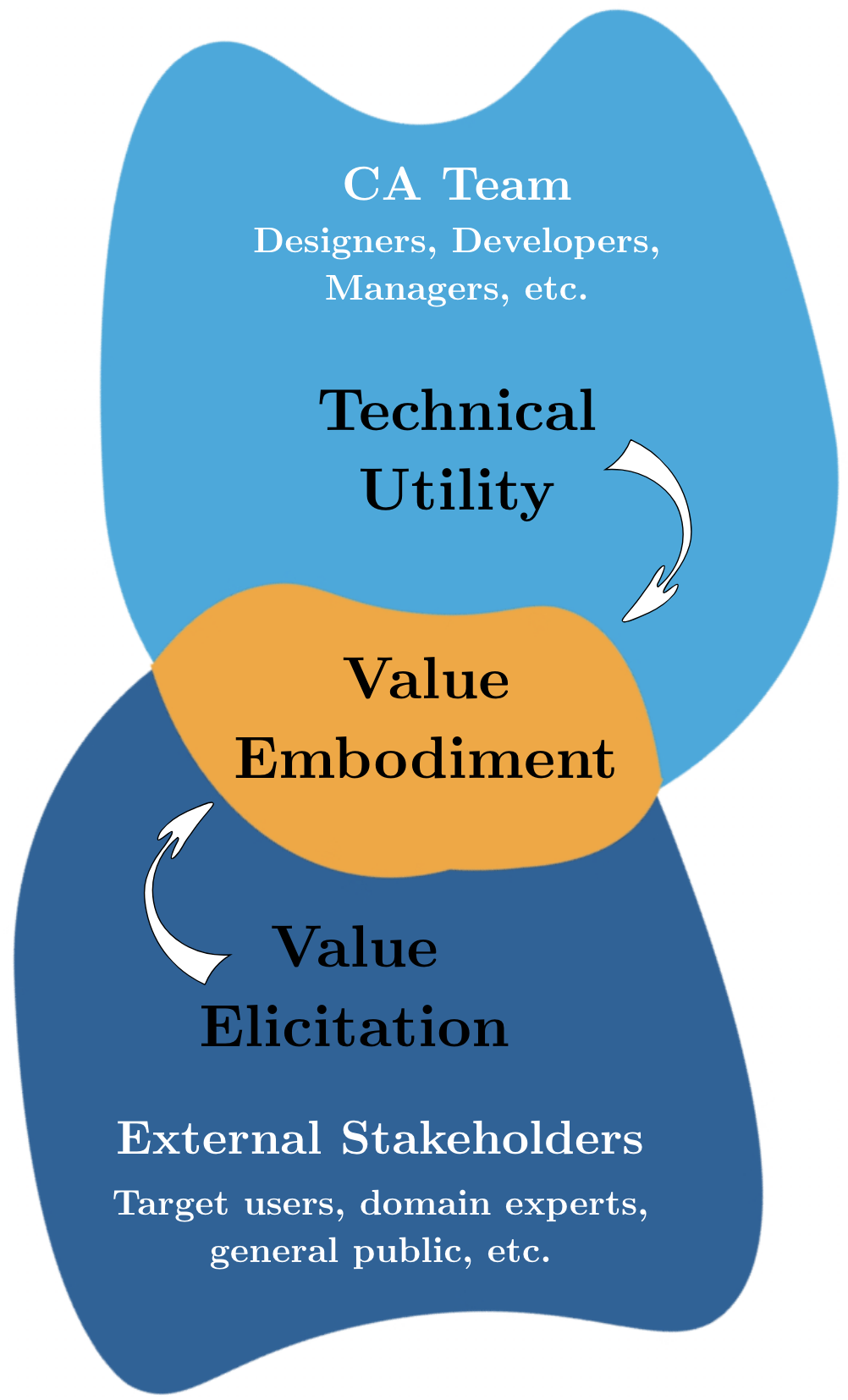}
    \caption{A model of the requirements needed to reach value embodiment which is framed as a collaborative effort between CA teams and external stakeholders. The values of external stakeholders need to be elicited, and collaboration outcomes need to have a technical utility in order for the CA team to contribute towards value embodiment.}
    \label{fig:barriers-model}
\end{figure}

\noindent Based on prior research conducted \citep{}, a model for co-designing value-sensitive CAs can now be conceptualised as shown in Figure \ref{fig:barriers-model}. There is little guidance regarding how to ensure that chosen values have been embedded into a technology \citep{manders2011}. \cite{poel2020} recommends focusing on ``embodied values" which are those that show up in the intermediate artefacts and prototypes created during the design of an AI system, as opposed to other forms of values that manifest before the design process begins or after it has concluded. In order to achieve value embodiment, there are requirements that arise from `external stakeholders' (including target users, domain experts, the general public, etc.) and from the `CA team' responsible for the CA (including designers, developers, managers, etc.). From the side of external stakeholders, value elicitation needs to take place in order to collect and understand their values. In order for the internal team to contribute towards value embodiment, the outcomes of the collaboration need to have technical utility - i.e. they need to be understandable and useful from a technical perspective in order to be accepted by practitioners and have the ability to drive CA implementation. 

\subsubsection{Value Elicitation}
Prior work has distinguished between using a top-down approach towards a technology's values where pre-defined values are obtained from an existing source, and a bottom-up approach where contextual values are obtained from the stakeholders of a technology \citep{vera2019}. A bottom-up approach is generally favored as it avoids several limitations of using pre-defined values, such as their lack of generalisability across definitions, groups of people, and contexts \citep{jakesch2022} and their abstract nature which makes them difficult to practically implement \citep{pommeranz2012}. The use of a bottom-up approach towards value elicitation has also been found to be effective in the context of AI systems \citep{verdiesen2022}, where creators' values can differ from those using the systems and affected by them \citep{jakesch2022}.

\subsubsection{Technical Utility}
For many co-design activities, it is unclear to technical practitioners how to use the outcomes produced. There is usually an extra effort of translation needed in order for outcomes to become useful from a technical perspective, in terms of practitioners being able to incorporate them into their workflows \citep{wong2022}. This contributes to technical practitioners having negative attitudes towards socio-technical design aspects \citep{manders2009, dindler2022} as they view these considerations as burdens or extra work that they must take on in addition to their usual technical responsibilities \citep{morley2021}. 

\bigskip
\noindent As mentioned earlier, the focus of the framework is on facilitating the co-design of boundary objects that correspond to different stages of the CA design process. These boundary objects are designed to elicit stakeholder values while holding technical utility for the CA team in order to enable the creation of value embodied prototypes. While the proposed interviews will explore value elicitation and technical utility, the proposed survey will focus on the resulting value embodiment of the prototypes produced.

\section{The VSCA Framework}
\label{sec:vsca-framework}

\subsection{Framework Structure}

\begin{figure}
    \centering
    \includegraphics[width=\textwidth]{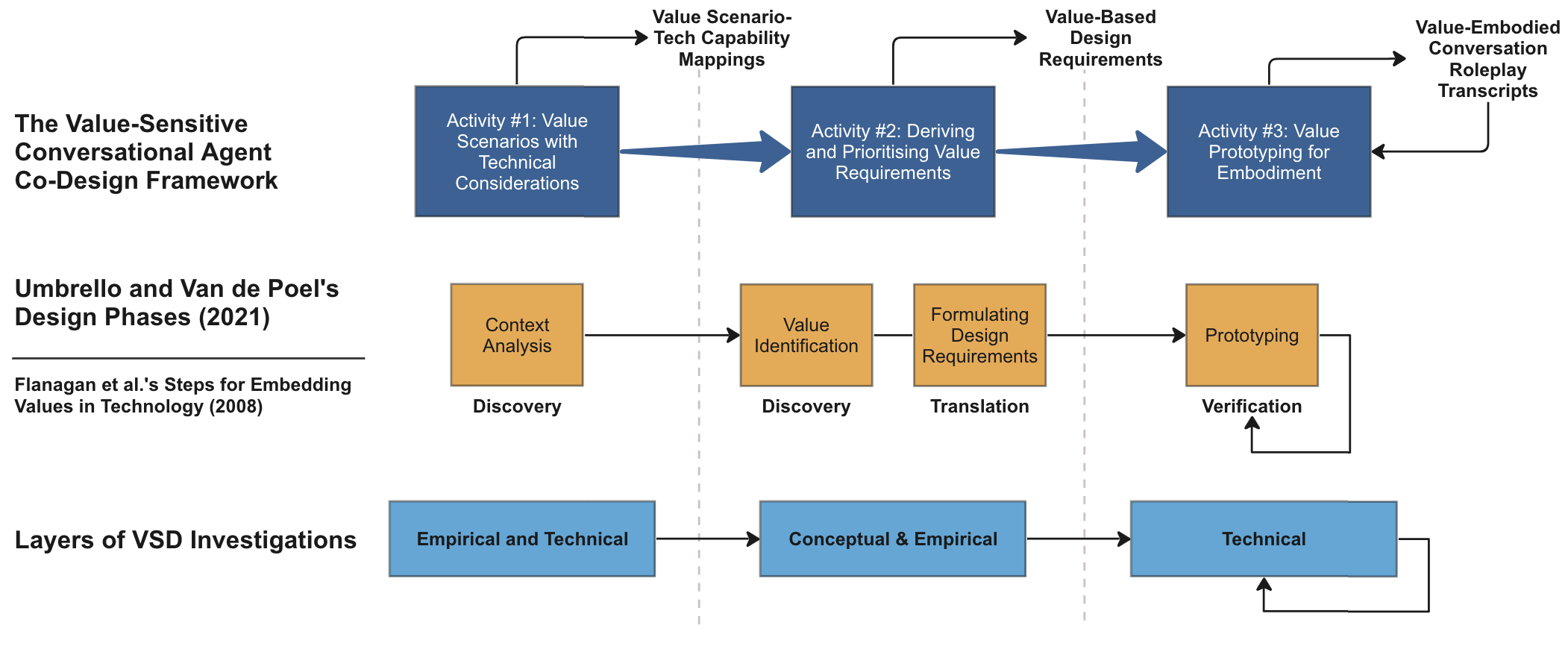}
    \caption{A diagram showing the activities and boundary objects included in the VSCA framework. These activities and objects are mapped to VSD \citep{friedman2006}, proposed steps for embedding values in technology \citep{flanagan2008}, and a conceptual framework for applying VSD to AI systems \citep{umbrello2018}.}
    \label{fig:framework-structure}
\end{figure}

\noindent The VSCA framework facilitates the creation of value embodied CA prototypes. It achieves this by guiding CA creators through three activities to co-design three boundary objects with their target users and other stakeholders. These boundary objects hold elicited stakeholder values and have a technical utility for practitioners to be able to use them to guide CA implementation. The activities and boundary objects are shown in Figure \ref{fig:framework-structure} which also maps the practical framework onto VSD \citep{friedman2006}, proposed steps for embedding values in technology \citep{flanagan2008}, and a conceptual framework for applying VSD to AI systems \citep{umbrello2018}.

\noindent The design activities which lead to the co-design of the three mentioned boundary objects are framed as value levers. Value levers are activities which facilitate discussions and actions around values during a project \citep{shilton2013}. These can be any activity, including gaining funding or simply working in an interdisciplinary team (ibid). Previous work has shown that design activities can funtion as value levers \citep{ahn2021} and that the use of value levers is effective in the context of AI systems \citep{varanasi2023, winecoff2022}. 

\bigskip
\noindent Each of the three activities has an identical structure which is based on the Value-Sensitive Action Reflection model \citep{yoo2013}. During each activity, participants are split into the CA team responsible for the CA (e.g. developers, designers, managers, etc.) and external stakeholders (e.g. target users, domain experts, the general public, etc.). Each team works on a task separately, then both teams come together to combine and consolidate their work into the given boundary object, while engaging in a value-lever discussion. The tasks that each team works on for each activity and the underlying literature that underpins each task are shown in Table \ref{tab:framework-activities}.

\begin{table}
    \caption{Each activity's CA team task, external stakeholder team task, value-lever discussion tasks, and resulting boundary objects.}
    \label{tab:framework-activities}
    \scriptsize
    \resizebox{\textwidth}{!}{
    \begin{tabular}{ccccc}
        \textbf{Activity} & \textbf{CA Team Task} & \makecell[c]{\textbf{External} \\ \textbf{Stakeholder} \\ \textbf{Team Task}} & \makecell[c]{\textbf{Value-Lever} \\ \textbf{Discussion} \\ \textbf{Task}} & \makecell[c]{\textbf{Resulting} \\ \textbf{Boundary Object}} \\
       \makecell[c]{\\ \textbf{Activity 1}} &  \makecell[c]{\\ Brainstorming technical \\ capabilities and \\ translating them \\ using metaphors\tablefootnote{\cite{yang2018, luria2018, alves2021}}} &  \makecell[c]{\\ Brainstorming value scenarios \\ using common CA design \\ decisions\tablefootnote{\cite{friedman2006, czeskis2010, vines2013, friedmanbook}} } &  \makecell[c]{\\ Mapping between \\ capabilities and \\ scenarios\tablefootnote{ \cite{ehsan2023, yildirim2023}}} &  \makecell[c]{\\ Value scenario \\ to technology capability \\ mappings}\\
        
        \makecell[c]{\\ \textbf{Activity 2}} &  \makecell[c]{\\ Deriving value-based \\ design requirements \tablefootnote{\cite{poel2013, spiekermann2021}}} &  \makecell[c]{\\ Value source analysis \tablefootnote{\cite{friedmanbook}}, \\ value categorising \tablefootnote{\cite{fuchsberger2012}} \\ and value prioritising \tablefootnote{\cite{umbrello2018}}} &  \makecell[c]{\\ Value dams and \\ flows \tablefootnote{\cite{borning2005, miller2007, denning2010}}} &  \makecell[c]{\\ Consensus of \\ value-based design \\ requirements}\\
        
       \makecell[c]{\\ \textbf{Activity 3}} &  \makecell[c]{\\ Conversation flowchart \\ design using \\ Natural Conversation \\ Framework \tablefootnote{\cite{moore2016, moore2018}}} &  \makecell[c]{\\ Conversation flowchart \\ design using \\ Natural Conversation \\ Framework\textsuperscript{9}} &  \makecell[c]{\\ Conversation \\ Roleplaying \tablefootnote{\cite{freier2008, friedmanbook}} \\ with discussion \\ prompts \tablefootnote{\cite{beinema2022}}} &  \makecell[c]{\\ Roleplay transcript, \\ conversation flowcharts, \\ discussion notes}\\
    \end{tabular}}
\end{table}

\subsection{Framework Operationalisation} 

\noindent The framework is operationalised using a design toolkit (shown at the beginning of the paper and broken down in the following section). The toolkit provides templates for each of the boundary objects that facilitates their creation and standardises their form. The goal is to provide a concrete and practical means for following the framework that CA creators can use in a design workshop setting to facilitate team activities and actualise the needed boundary objects. The toolkit is composed of a guidebook, a set of canvases for the team activities and the boundary objects, a set of blank cards to be filled out for the first boundary object, and information sheets that provide helpful information during different activities in order to keep participants' ideas focused and relevant. 

\subsection{Framework Activities}

\subsubsection{Activity \#1 - Value Scenarios with Technical Considerations} 

\noindent The objective of Activity 1 is to conduct a contextual analysis in order to map stakeholders’ needs and values to available technological capabilities and opportunities. CA team members work on brainstorming technological capabilities available to them and are encouraged to use metaphors to explain them to non-technical stakeholders as needed. Metaphors can help people understand and design conversational agents through anthropomorphism \citep{luria2018} and can help them brainstorm new ideas revolving around new technologies \citep{alves2021}. These capabilities can include the ability to collect data through sensors or questions, the ability to infer different types of data from each other, the ability to do things on users' behalf using APIs, and so on. 

\bigskip
\noindent Meanwhile, external stakeholders create value scenarios \citep{friedman2006} to describe different use-cases for the envisioned CA and root them in values important to them. In this case, the value scenarios are used for value elicitation, as well as technological framing ~[p. 88]\citep{friedmanbook}. In order to focus on realistic scenarios, the external stakeholder team are also provided with a list of common CA design decisions, such as the modality (e.g. text or voice), the channel it will be presented through (e.g. website, social media app, etc.), its personality, and so on. These design decisions can help the team envision the conversational agent they need more realistically and understand what kinds of decisions they need to address through their scenarios. The goal of this task is to begin eliciting stakeholder values in an indirect way while also providing envisioned use-cases for the CA team. Several experts have recommended that directly asking stakeholders about their values is not always the most effective approach, and that it is best to indirectly guide them towards understanding their values \citep{}, which might be unconscious or unfelt \citep{cockton2005, cockton2006}, through the use of envisioning and other design activities \citep{lim2022}. 

\bigskip
\noindent After the team-based tasks, the group will come together to consolidate their outputs. The internal team will present their identified technology capabilities and opportunities with the help of the metaphors available to them when needed, while the external stakeholders will present their value scenarios. Each team will have written their ideas on a set of blank cards. During this task, they will match the cards together given which capabilities or opportunities can address which needs and values in the given scenarios, and map them onto a canvas provided. Afterwards, unaddressed scenarios and unutilised capabilities can be discussed. 

\begin{figure}
    \centering
    \includegraphics[width=\textwidth]{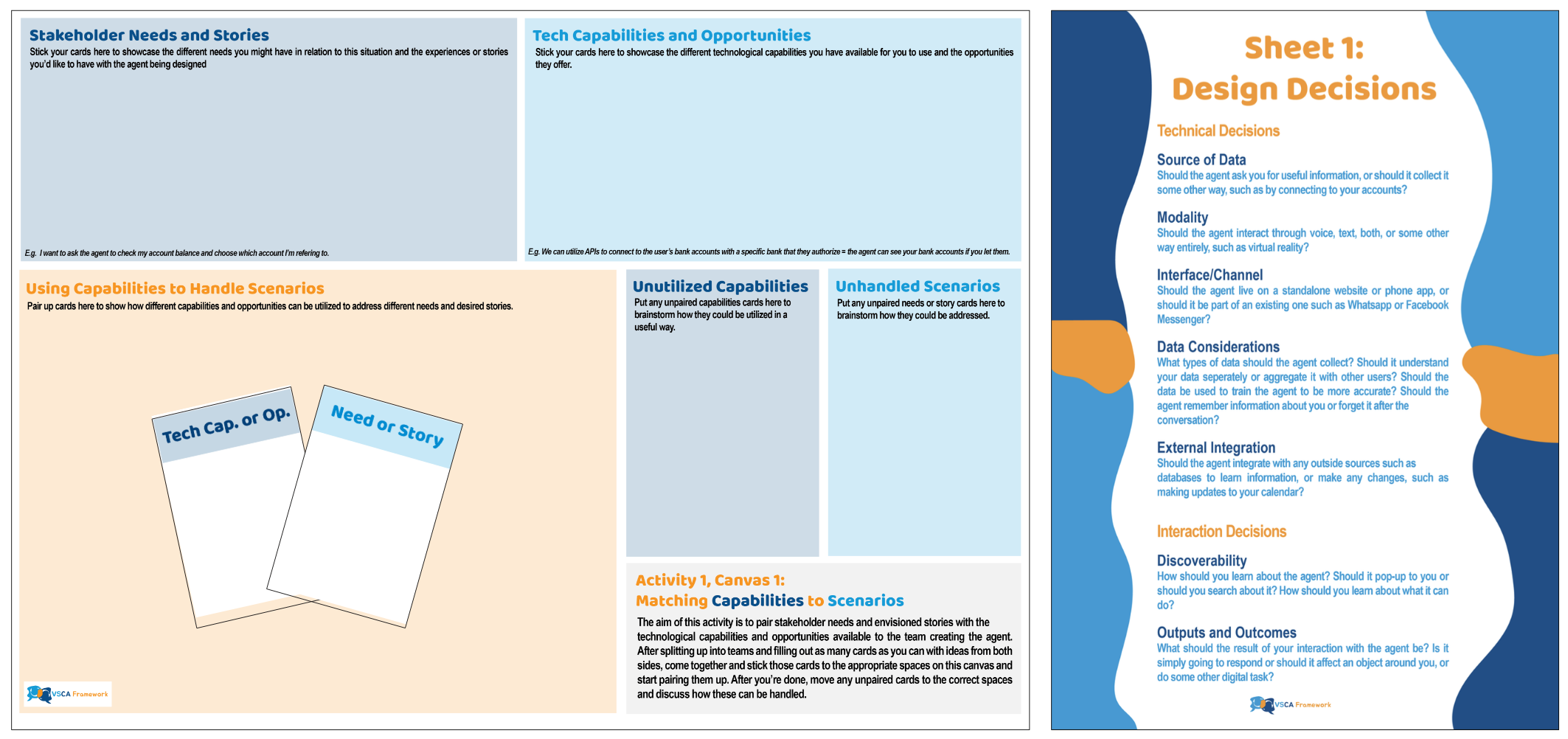}
    \caption{Activity \#1 toolkit components: Blank cards for brainstorming technological capabilities and value scenarios; a canvas with allocated spaces for capabilities, scenarios, matched pairs, unutilised capabilities, and unhandled scenarios; and an information sheet with common CA design decisions to guide scenario brainstorming.}
    \label{fig:activity-1}
\end{figure}

\bigskip
\noindent The activity focuses on mapping value-based scenarios originating from stakeholders to technology capabilities and opportunities provided by the design team. The goal is to focus specific technology-based decisions on addressing needs and values and avoiding the tendency of ``fitting people to technologies" \citep{vines2013} or having a ``solution in search of a need" \citep{calvo2017}. This consideration of human-side and technology-side aspects and bridging between them also has parallels in: (i) the socio-technical gap described by \cite{ehsan2023}, (ii) \cite{yildirim2023}’s findings of how AI practitioners work to ``match AI capabilities with human needs”, and (iii) the approach taken by the Conversational Design Institute (CDI)’s in their Conversation Design Canvas\footnote{The canvas is accessible at: https://www.conversationdesigninstitute.com/communications/cd-tool-kit}. The key differences in this instance is the focus on values, the facilitation of communication through the use of metaphors, and the scoping of envisioned scenarios by grounding them in common CA design decisions. Figure \ref{fig:activity-1} shows the toolkit components used during this activity.

\subsubsection{Activity \#2 - Deriving and Prioritising Value Requirements}

\noindent The objective of Activity 2 is to extract and categorize values from the [value scenario-technology capability] mappings created during Activity 1 and convert them into value-based design requirements for the envisioned CA. For the initial task conducted in two individual groups (CA team vs. external stakeholders), the external stakeholders will engage in value source analysis ~[p. 62]\citep{friedmanbook} and categorisation based on the value scenarios produced during the previous workshop, while the CA team begin translating the [value scenarios-technology capability] mappings into value-based design requirements. 

\bigskip
\noindent External stakeholders will begin working with the [value scenario-technology capability mappings] they created during the previous workshop. Firstly, they will identify a set of values underlying the different mappings and identify which originate from the CA team and which originate from external stakeholders. The separation of CA team values from external stakeholder values in the value elicitation process forms a kind of value source analysis. This separation is recommended in VSD as a method for more clearly identifying and resolving tensions ~[p. 68]\citep{friedmanbook}. After making the values explicit, they will categorise them based on the Values in Action (ViA) approach \citep{fuchsberger2012} which builds upon the Theory of Consumption Values (TCV) \citep{hedman2010}. The ViA approach considers both functional/technical values relating to the usability and utility of the technology and the hedonic values of users relating to their emotions and experiences \citep{fuchsberger2012}. The approach builds upon TCV and includes both interpersonal values along with other values relating to usability, user experience and user acceptance. The approach allows for taxonomising values and understanding which value classes are of most relevant in different contexts in order to prioritise those. This categorisation helps to establish a hierarchy of values which can help rank values and resolve value tensions \citep{manders2011}, de-bias elicited values \citep{umbrello2018}, and can help technical practitioners understand which values serve which aspects of the CA. 

\bigskip
\noindent Meanwhile, the CA team will write out value-based design requirements based on the [value scenario-technology capability] mappings. A number of previous works have conceptually derived design requirements for AI systems from sets of ethical principles \citep{kroes2020, poel2013, ieee7000}. As such, internal team members derive value-based design requirements that mention the use-case targeted, the technological capability utilitsed, and the underlying values involved. They will be given a few examples to help them understand the process. 

\begin{figure}
    \centering
    \includegraphics[width=\textwidth]{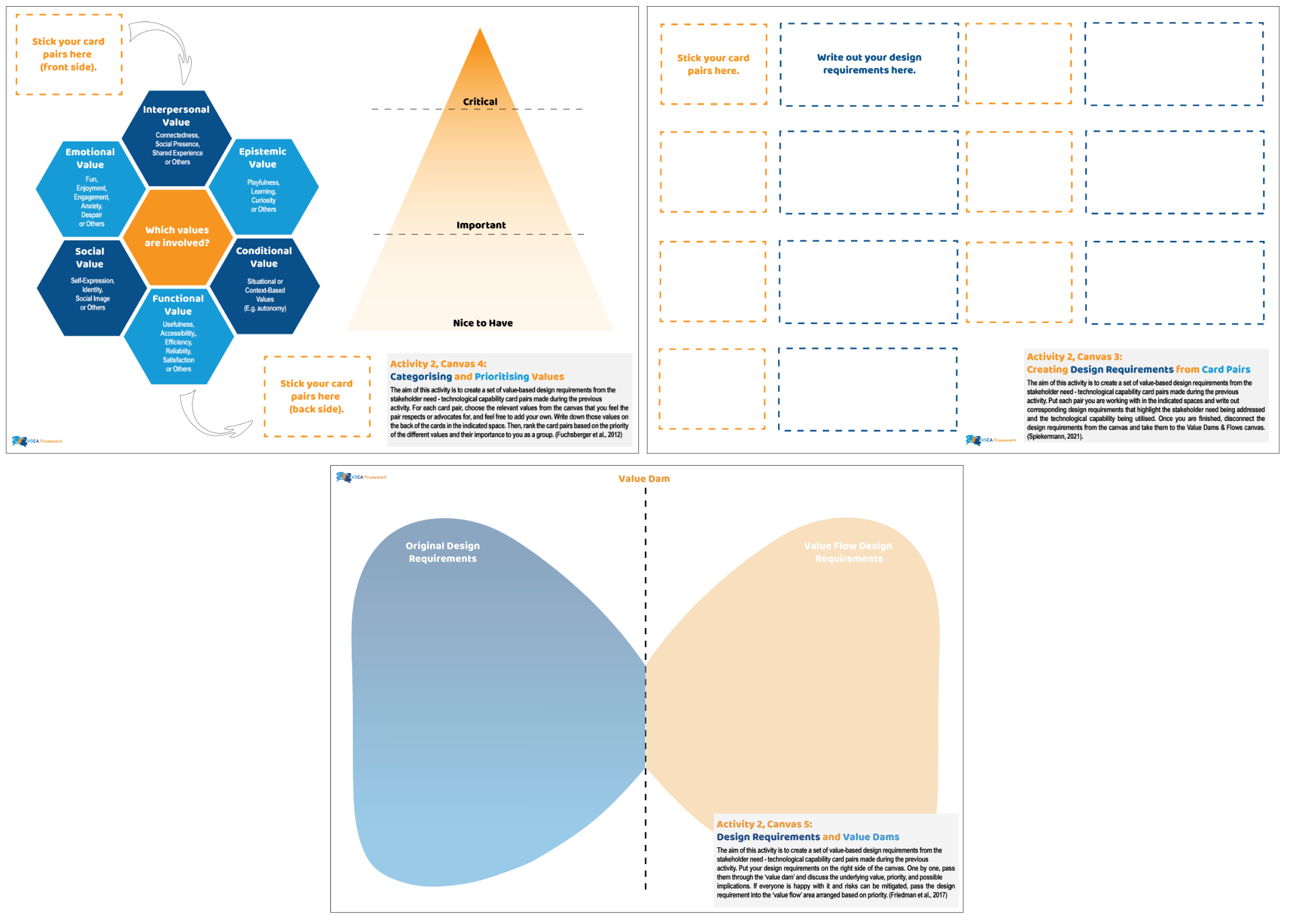}
    \caption{Activity \#2 toolkit components: Three canvases for the three tasks involved. The top left canvas is used by the external stakeholders to conduct value source analysis, categorisation and prioritisation using the backs of the cards used during Activity \#1 which have spaces for writing values and priority levels. The top right canvas is used by the CA team to write value-based design requirements for different pairs of [technology capability - value scenario] cards. The bottom canvas is for the value dams and flows exercise where initial design requirements are placed on the right and then passed through the dam to the left after modifications are made and a consensus is reached.}
    \label{fig:activity-2}
\end{figure}

\bigskip
\noindent Both groups will then come together and share their outcomes. External stakeholders will present their categorised values and CA team members will present their initial derived design requirements. The group can then derive extra design requirements based on the categorised values and modify existing design requirements to explicitly mention the underlying values to be preserved in each case. Value Dams and Flows are then used, which have proven useful in other value-oriented studies for value-based analysis and selection of criteria/features/etc. \citep{borning2005, denning2010, miller2007, czeskis2010}. Unlike other works, Value Dams and Flows are used here in a more participatory approach where the group collectively discusses and decides if any tensions are present and which requirements can `pass the dam' and `flow' through. The resulting value-embodied artifact takes the form of a set of value-based design requirements derived from the value scenarios created during the previous workshop and the elicited underlying categorised values. Figure \ref{fig:activity-2} shows the toolkit components used during this activity.

\subsubsection{Activity \#3 - Value Prototyping for Embodiment}

\noindent The objective of Activity 3 is to create a conversational prototype based on the value scenarios and design requirements produced in the previous two activities. For the initial task conducted in two individual groups (CA team vs. external stakeholders), both teams will separately create conversation dialogue flows for any number of value scenarios outlined in previous workshops. In order to aid them in doing so, they will be provided with conversation patterns from the Natural Conversation Framework \citep{moore2016, moore2018}. The Natural Conversation Framework (NCF) offers an extensive set of simple conversational user interaction patterns. The NCF's interaction model consists of a set of 15 patterns or sequences which each have several possible expansions. One of the particularly powerful features of the NCF is its compatibility with the design life-cycle, with work already done to consolidate the two \citep{moore2019}. The goal of providing extracts from the NCF is to ensure participants' conversational flows adhere to standards and are created in a form that is directly usable during implementation.

\begin{figure}
    \centering
    \includegraphics[width=0.8\textwidth]{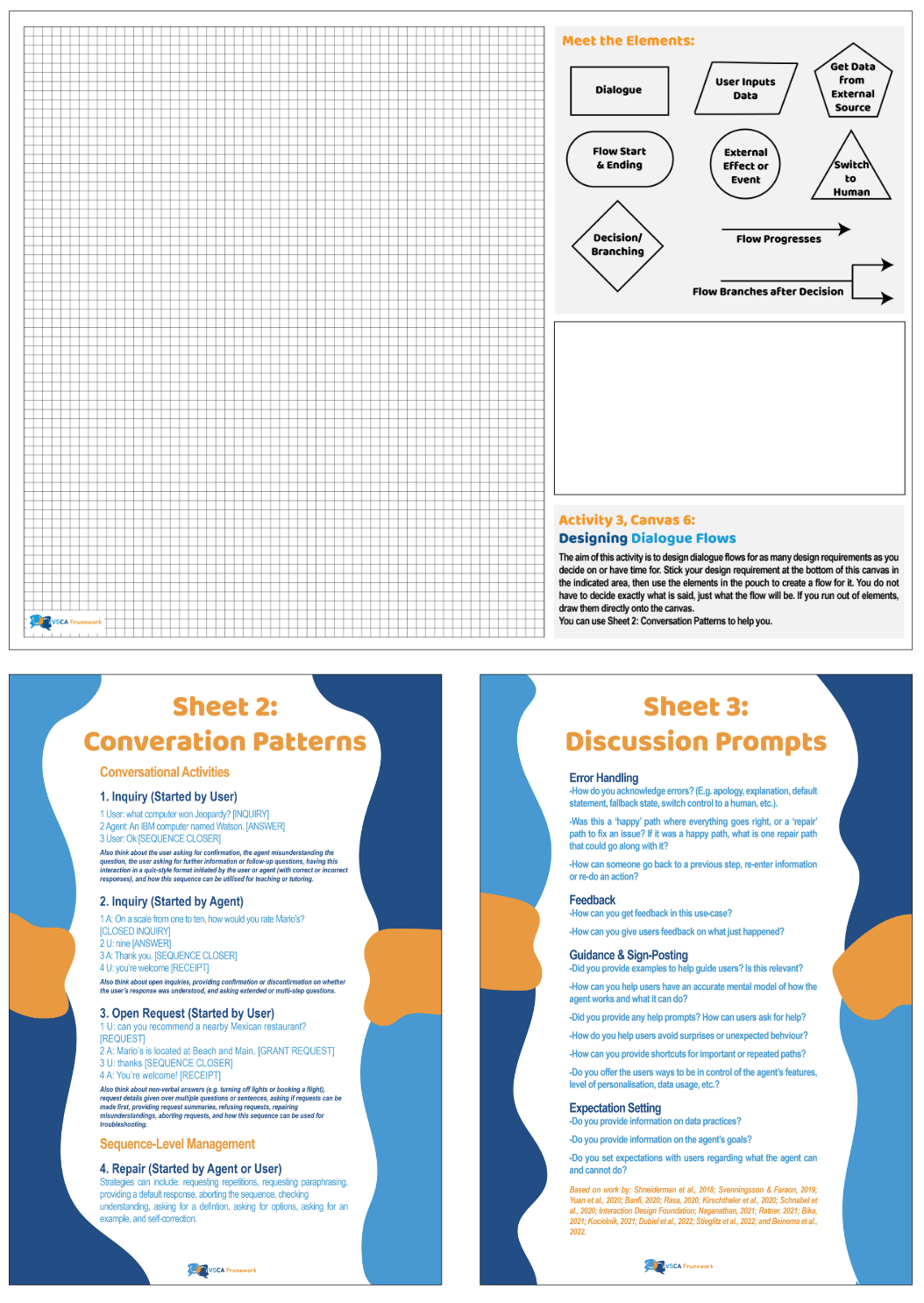}
    \caption{Activity \#3 toolkit components: a canvas for drawing conversational flowcharts and information sheets for NCF patterns and prompts to guide discussions during conversation roleplaying.}
    \label{fig:activity-3}
\end{figure}

\bigskip
\noindent As the two groups come together, they will then roleplay CA team members' and external stakeholders' different versions of a scenario, which is a very commonly used technique in CA co-design \citep{garg2020, easton2019, candello2020, lee2017, weber2021, luria2020, lopatovska2022}. They can then discuss differences in content, tone of voice, agent personality, and so on using a set of discussion prompts provided. The transcripts of these roleplay exercises form a prototype which embodies the learnings, elicited values, and outcomes of the workshops in a form usable by technical practitioners. The use of value-oriented mockups or prototypes is another method described by VSD ~[p. 62]\citep{friedmanbook} as a way of contextualizing the technology under investigation in real-world settings, and has been used in value-oriented investigations \citep{yoo2013}. While they can take on several forms and mediums (e.g. \cite{denning2010} use tangible prototypes, while \cite{woelfer2009} use video prototypes), value-oriented Wizard of Oz prototypes enabled through role-playing have specifically been used before to understand the implications of personified agents \citep{freier2008}. The resulting value-embodied boundary object takes the form of the recorded roleplay transcripts, the notes of resulting discussions and changes, and the conversational flows. Figure \ref{fig:activity-3} shows the toolkit components used during this activity.

\section{Proposed Evaluation}

\subsection{Design Workshop Case Study and Semi-Structured Interviews}
\noindent The framework and toolkit will be used in a design workshop setting with a team building a CA and their stakeholders. The three activities will be conducted during the workshop to produce the three boundary objects needed. Qualitative data will be collected in the form of researcher observations during the workshop, and semi-structured interviews with participants after the workshop. Interviews with the CA team will focus on their perceived technical utility, including how they envision using the boundary objects in their workflows and how the framework fits in with their existing practices. Interviews with external stakeholders will examine their perceptions on value elicitation throughout the process and the value embodiment of the outcomes produced.

\subsection{Survey to Assess Value Embodiment of Prototype Produced}
\noindent Following the workshop, the roleplay transcript will be converted into a CA script. An independent conversation designer will be asked to produce another CA script given the same brief given to workshop participants and one of the use-cases created during the workshop, but without following the framework. The goal is to mimic current industry practices and provide a control condition. Both scripts will then be presented to the CA team's target users through a survey. Survey takers will be asked to rate the extent with which they feel each script embodies the intended values elicited from the workshops, as well as their own values. This will allow for an evaluation of both scripts based on the degree of perceived alignment between embodied values and intended values that surfaced during the workshop \citep{poel2020}, as well as the degree of perceibed value alignment of each script with participants' (target users') values.

\section{Discussion}
\subsection{Strengths of the Framework}
The VSCA framework offers a concrete and practical approach towards ensuring CAs are co-designed with relevant stakeholders and embody their values. It builds upon largely conceptual work in the space of VSD and harnesses the power of frameworks over other forms of interventions to operationalise key objectives. The framework is grounded in existing literature and addresses a model of factors influencing the co-design of value-sensitive CAs, which is based on previous research. It abstracts away different modalities of CAs (e.g. voice versus text) and aims to generalise beyond specific applications or domains. It also ensures that the outcomes produced are focused, relevant, and aligned with existing standards and development practices. It leverages technical utility and stakeholder values to achieve value embodiment, and its impact across these three aspects will be explored empirically using a mixed-method approach.

\subsection{Limitations and Future Work}
Exploring value-related aspects in CA design is an immensely broad topic that is difficult to sufficiently cover in the scope of one project. As such, this project only scratches the surface and leaves a number of stones unturned for future work to explore. Firstly, it must be acknowledged that the context of this work is Euro-centric in terms of all the studies' participants and the values they expressed. It will also predictably suffer from a limited number of participants in the design workshops and subsequent interviews given the time commitment involved, although the richness of the qualitative data gathered aims to offset this limitation. In terms of working with values in general, although some strategies are in place to address these aspects, the framework cannot claim to fully address the role of value conflicts and trade-offs \citep{vermaas2014}, power hierarchies between participants \citep{park2022}, and a number of limitations regarding value-sensitive design \citep{friedman2021, umbrello2021a} and working with embodied values \citep{poel2014}.

\bigskip
\noindent In terms of future work, it should aim to be more intercultural, intersectional and tangential. Additionally, one of the significant struggles faced throughout this project was the lack of a metric or measurement tool for value-sensitivity or value embodiment. This work and others like it would benefit greatly from the development of such a standard measure to overcome several of the difficulties of working with values, especially their subjective, abstract, and ephemeral nature. There is also still much more to explore regarding the relationship between values and emotions in the context of CAs, and while existing work has begun to explore the role of emotions \citep{gornemann2022}, a formal link between both has not been established as of yet. Emotions could provide a more concrete and measurable indicator of value alignment or violation. Finally, future work also needs to examine large language models and how the identified requirements, and value embodiment in general, are affected by these models and people's perception and use of them. 

\section{Conclusion}
This paper has introduced the Value-Sensitive Conversational Agent (VSCA) Framework. The framework was built upon several strands of discourse and literature, and was grounded in the objective of providing a process for co-designing value-sensitive CAs. The framework enables the co-design of three boundary objects between CA teams and their stakeholders to collect stakeholder values and provide technical utility to drive CA implementation. The boundary objects facilitate the creation of value embodied CA prototypes. Upcoming work will evaluate the framework using a mixed-method approach where value elicitation and technical utility are explored qualitatively through semi-structured interviews with participants who use the framework during a design workshop, and value embodiment will be assessed quantitatively via a survey with the designed CA's target users using the prototype created during the workshop.

\bigskip
\noindent This space is still rapidly emerging and while it is exciting to work at the intersection of CAs and design, it can also be overwhelming to keep up with the constant developments unfolding. Only as a community can we hope to impact both academic research and industry-based practices. By sharing the framework developed and the planned evaluation, we hope to inspire future work across the community regarding aligning CAs with stakeholders' and societies' values, and make their collaborative design more practical, systematic, and widespread.








\Urlmuskip=0mu plus 1mu
\bibliographystyle{ACM-Reference-Format}
\bibliography{main}


\begin{thebibliography}{120}


\ifx \showCODEN    \undefined \def \showCODEN     #1{\unskip}     \fi
\ifx \showDOI      \undefined \def \showDOI       #1{#1}\fi
\ifx \showISBNx    \undefined \def \showISBNx     #1{\unskip}     \fi
\ifx \showISBNxiii \undefined \def \showISBNxiii  #1{\unskip}     \fi
\ifx \showISSN     \undefined \def \showISSN      #1{\unskip}     \fi
\ifx \showLCCN     \undefined \def \showLCCN      #1{\unskip}     \fi
\ifx \shownote     \undefined \def \shownote      #1{#1}          \fi
\ifx \showarticletitle \undefined \def \showarticletitle #1{#1}   \fi
\ifx \showURL      \undefined \def \showURL       {\relax}        \fi
\providecommand\bibfield[2]{#2}
\providecommand\bibinfo[2]{#2}
\providecommand\natexlab[1]{#1}
\providecommand\showeprint[2][]{arXiv:#2}

\bibitem[Abaza(2021)]%
        {abaza2021}
\bibfield{author}{\bibinfo{person}{A. Abaza}.} \bibinfo{year}{2021}\natexlab{}.
\newblock \bibinfo{title}{MLOps: Why is it the Most Important Technology in the
  Age of AI?}
\newblock
\newblock
\urldef\tempurl%
\url{https://www.synapse-analytics.io/post/mlops-why-is-it-the-most-important-technology-in-the-age-of-ai}
\showURL{%
\tempurl}


\bibitem[Ahn et~al\mbox{.}(2021)]%
        {ahn2021}
\bibfield{author}{\bibinfo{person}{J. Ahn}, \bibinfo{person}{F. Campos},
  \bibinfo{person}{H. Nguyen}, \bibinfo{person}{M. Hays}, {and}
  \bibinfo{person}{J. Morrison}.} \bibinfo{year}{2021}\natexlab{}.
\newblock \showarticletitle{Co-Designing for Privacy, Transparency, and Trust
  in K-12 Learning Analytics}. \bibinfo{publisher}{Proceedings of the
  International Learning Analytics and Knowledge Conference (LAK)}.
\newblock
\urldef\tempurl%
\url{https://doi.org/10.1145/3448139.3448145}
\showDOI{\tempurl}


\bibitem[Alves-Oliveira et~al\mbox{.}(2021)]%
        {alves2021}
\bibfield{author}{\bibinfo{person}{P. Alves-Oliveira}, \bibinfo{person}{M.
  Lupetti}, \bibinfo{person}{M. Luria}, \bibinfo{person}{D. Löffler},
  \bibinfo{person}{M. Gamboa}, \bibinfo{person}{L. Albaugh},
  \bibinfo{person}{W. Kamino}, \bibinfo{person}{A. Ostrowski},
  \bibinfo{person}{D. Puljiz}, \bibinfo{person}{P. Reynolds-Cuéllar},
  \bibinfo{person}{M. Scheunemann}, \bibinfo{person}{M. Suguitan}, {and}
  \bibinfo{person}{D. Lockton}.} \bibinfo{year}{2021}\natexlab{}.
\newblock \showarticletitle{Collection of Metaphors for Human-Robot
  Interaction}. In \bibinfo{booktitle}{\emph{Proceedings of the Designing
  Interactive Systems Conference (DIS)}}.
\newblock


\bibitem[Arrighi and Mougenot(2019)]%
        {arrighi2019}
\bibfield{author}{\bibinfo{person}{PA. Arrighi} {and} \bibinfo{person}{C.
  Mougenot}.} \bibinfo{year}{2019}\natexlab{}.
\newblock \showarticletitle{Towards user empowerment in product design: a mixed
  reality tool for interactive virtual prototyping}.
\newblock \bibinfo{journal}{\emph{J Intell Manuf}}  \bibinfo{volume}{30}
  (\bibinfo{year}{2019}), \bibinfo{pages}{743--754}.
\newblock


\bibitem[Axelsson et~al\mbox{.}(2021)]%
        {axelsson2021}
\bibfield{author}{\bibinfo{person}{Minja Axelsson}, \bibinfo{person}{Raquel
  Oliveira}, \bibinfo{person}{Mattia Racca}, {and} \bibinfo{person}{Ville
  Kyrki}.} \bibinfo{year}{2021}\natexlab{}.
\newblock \showarticletitle{Social Robot Co-Design Canvases: A Participatory
  Design Framework}.
\newblock \bibinfo{journal}{\emph{J. Hum.-Robot Interact.}}
  \bibinfo{volume}{11}, \bibinfo{number}{1}, Article \bibinfo{articleno}{3}
  (\bibinfo{date}{oct} \bibinfo{year}{2021}), \bibinfo{numpages}{39}~pages.
\newblock
\urldef\tempurl%
\url{https://doi.org/10.1145/3472225}
\showDOI{\tempurl}


\bibitem[Beinema et~al\mbox{.}(2022)]%
        {beinema2022}
\bibfield{author}{\bibinfo{person}{T. Beinema}, \bibinfo{person}{H. Akker},
  \bibinfo{person}{H. Hermens}, {and} \bibinfo{person}{L. van Velsen}.}
  \bibinfo{year}{2022}\natexlab{}.
\newblock \showarticletitle{What to Discuss? — A Blueprint Topic Model for
  Health Coaching Dialogues With Conversational Agents}.
\newblock \bibinfo{journal}{\emph{International Journal of Human–Computer
  Interaction}} (\bibinfo{year}{2022}).
\newblock


\bibitem[Borning et~al\mbox{.}(2005)]%
        {borning2005}
\bibfield{author}{\bibinfo{person}{A. Borning}, \bibinfo{person}{B. Friedman},
  \bibinfo{person}{J. Davis}, {and} \bibinfo{person}{P. Lin}.}
  \bibinfo{year}{2005}\natexlab{}.
\newblock \showarticletitle{Informing Public Deliberation: Value Sensitive
  Design of Indicators for a Large-Scale Urban Simulation}.
  \bibinfo{publisher}{ECSCW 2005}.
\newblock


\bibitem[Boujut and Blanco(2003)]%
        {boujut2003}
\bibfield{author}{\bibinfo{person}{J.F. Boujut} {and} \bibinfo{person}{E.
  Blanco}.} \bibinfo{year}{2003}\natexlab{}.
\newblock \showarticletitle{Intermediary Objects as a Means to Foster
  Co-operation in Engineering Design}.
\newblock \bibinfo{journal}{\emph{Computer Supported Cooperative Work (CSCW)}}
  \bibinfo{volume}{12} (\bibinfo{year}{2003}), \bibinfo{pages}{205--219}.
\newblock


\bibitem[Byrd(2020)]%
        {byrd2020}
\bibfield{author}{\bibinfo{person}{L. Byrd}.} \bibinfo{year}{2020}\natexlab{}.
\newblock \bibinfo{title}{Tech Inclusion: How culture may enhance
  conversational AI technology}.
\newblock
\newblock
\urldef\tempurl%
\url{https://uxmag.com/articles/tech-inclusion-how-culture-may-enhance-conversational-ai-technology}
\showURL{%
\tempurl}


\bibitem[Cabitza and Simone(2012)]%
        {cabitza2012}
\bibfield{author}{\bibinfo{person}{F. Cabitza} {and} \bibinfo{person}{C.
  Simone}.} \bibinfo{year}{2012}\natexlab{}.
\newblock \showarticletitle{Affording Mechanisms: An Integrated View of
  Coordination and Knowledge Management}.
\newblock \bibinfo{journal}{\emph{Computer Supported Cooperative Work (CSCW)}}
  \bibinfo{volume}{21} (\bibinfo{year}{2012}), \bibinfo{pages}{227--260}.
\newblock


\bibitem[Calvo and Peters(2017)]%
        {calvo2017}
\bibfield{author}{\bibinfo{person}{R. Calvo} {and} \bibinfo{person}{D.
  Peters}.} \bibinfo{year}{2017}\natexlab{}.
\newblock \bibinfo{booktitle}{\emph{Positive Computing}}.
\newblock \bibinfo{publisher}{MIT Press}.
\newblock


\bibitem[Candello et~al\mbox{.}(2020)]%
        {candello2020}
\bibfield{author}{\bibinfo{person}{H. Candello}, \bibinfo{person}{M.
  Pichiliani}, \bibinfo{person}{C. Pinhanez}, \bibinfo{person}{S. Vidon}, {and}
  \bibinfo{person}{M. Wessel}.} \bibinfo{year}{2020}\natexlab{}.
\newblock \showarticletitle{Co-Designing a Conversational Interactive Exhibit
  for Children}. \bibinfo{publisher}{Proceedings of the ACM Interaction Design
  and Children Conference: Extended Abstracts}.
\newblock


\bibitem[Chen et~al\mbox{.}(2023)]%
        {chen2023}
\bibfield{author}{\bibinfo{person}{X. Chen}, \bibinfo{person}{J. Burke},
  \bibinfo{person}{R. Du}, \bibinfo{person}{M. Hong}, \bibinfo{person}{J.
  Jacobs}, \bibinfo{person}{P. Laban}, \bibinfo{person}{D. Li},
  \bibinfo{person}{N. Peng}, \bibinfo{person}{K. Willis}, \bibinfo{person}{C.
  Wu}, {and} \bibinfo{person}{B. Zhou}.} \bibinfo{year}{2023}\natexlab{}.
\newblock \showarticletitle{Next Steps for Human-Centered Generative AI: A
  Technical Perspective}.
\newblock \bibinfo{journal}{\emph{Computing Research Repository (CoRR)}}
  (\bibinfo{year}{2023}).
\newblock
\urldef\tempurl%
\url{https://arxiv.org/abs/2306.15774}
\showURL{%
\tempurl}


\bibitem[Chen et~al\mbox{.}(2020)]%
        {chen2020}
\bibfield{author}{\bibinfo{person}{Z. Chen}, \bibinfo{person}{Y. Lu}, {and}
  \bibinfo{person}{A. Nieminen, M.and~Lucero}.}
  \bibinfo{year}{2020}\natexlab{}.
\newblock \showarticletitle{Creating a Chatbot for and with Migrants: Chatbot
  Personality Drives Co-Design Activities}. In
  \bibinfo{booktitle}{\emph{Proceedings of the ACM Designing Interactive
  Systems Conference}}.
\newblock


\bibitem[Cherubini(2017)]%
        {cherubini2017}
\bibfield{author}{\bibinfo{person}{M. Cherubini}.}
  \bibinfo{year}{2017}\natexlab{}.
\newblock \bibinfo{title}{Ethical Autonomous Algorithms}.
\newblock
\newblock
\urldef\tempurl%
\url{https://medium.com/@mchrbn/ethical-autonomous-algorithms-5ad07c311bcc}
\showURL{%
\tempurl}


\bibitem[Christian(2020)]%
        {christian2020}
\bibfield{author}{\bibinfo{person}{B. Christian}.}
  \bibinfo{year}{2020}\natexlab{}.
\newblock \bibinfo{booktitle}{\emph{The Alignment Problem – Machine Learning
  and Human Values}}.
\newblock \bibinfo{publisher}{W. W. Norton \& Company}.
\newblock


\bibitem[Cockton(2005)]%
        {cockton2005}
\bibfield{author}{\bibinfo{person}{Gilbert Cockton}.}
  \bibinfo{year}{2005}\natexlab{}.
\newblock \showarticletitle{A Development Framework for Value-Centred Design}.
  \bibinfo{publisher}{CHI Extended Abstracts on Human Factors in Computing
  Systems}.
\newblock


\bibitem[Cockton(2006)]%
        {cockton2006}
\bibfield{author}{\bibinfo{person}{Gilbert Cockton}.}
  \bibinfo{year}{2006}\natexlab{}.
\newblock \showarticletitle{Designing Worth is Worth Designing}.
  \bibinfo{publisher}{Proceedings of the Nordic Conference on Human-Computer
  Interaction}.
\newblock


\bibitem[Crawford and Calo(2016)]%
        {crawford2016}
\bibfield{author}{\bibinfo{person}{K. Crawford} {and} \bibinfo{person}{R.
  Calo}.} \bibinfo{year}{2016}\natexlab{}.
\newblock \showarticletitle{There is a blind spot in AI research}.
\newblock \bibinfo{journal}{\emph{Nature}}  \bibinfo{volume}{538}
  (\bibinfo{year}{2016}), \bibinfo{pages}{311--313}.
\newblock


\bibitem[Custis(2021)]%
        {custis2021}
\bibfield{author}{\bibinfo{person}{C. Custis}.}
  \bibinfo{year}{2021}\natexlab{}.
\newblock \bibinfo{title}{Operationalizing AI Ethics through Documentation:
  ABOUT ML in 2020 and Beyond}.
\newblock
\newblock
\urldef\tempurl%
\url{https://partnershiponai.org/about-ml-2021/}
\showURL{%
\tempurl}


\bibitem[Czeskis et~al\mbox{.}(2010)]%
        {czeskis2010}
\bibfield{author}{\bibinfo{person}{Alexei Czeskis}, \bibinfo{person}{Ivayla
  Dermendjieva}, \bibinfo{person}{Hussein Yapit}, \bibinfo{person}{Alan
  Borning}, \bibinfo{person}{Batya Friedman}, \bibinfo{person}{Brian Gill},
  {and} \bibinfo{person}{Tadayoshi Kohno}.} \bibinfo{year}{2010}\natexlab{}.
\newblock \showarticletitle{Parenting from the Pocket: Value Tensions and
  Technical Directions for Secure and Private Parent-Teen Mobile Safety}.
  \bibinfo{publisher}{Proceedings of the Sixth Symposium on Usable Privacy and
  Security}.
\newblock


\bibitem[Dantec et~al\mbox{.}(2009)]%
        {dantec2009}
\bibfield{author}{\bibinfo{person}{C. Dantec}, \bibinfo{person}{E. Poole},
  {and} \bibinfo{person}{S. Wyche}.} \bibinfo{year}{2009}\natexlab{}.
\newblock \showarticletitle{Values As lived experience: evolving value
  sensitive design in support of value discovery}.
  \bibinfo{publisher}{Proceedings of the SIGCHI Conference on Human Factors in
  Computing Systems}.
\newblock


\bibitem[Denning et~al\mbox{.}(2010)]%
        {denning2010}
\bibfield{author}{\bibinfo{person}{Tamara Denning}, \bibinfo{person}{Alan
  Borning}, \bibinfo{person}{Batya Friedman}, \bibinfo{person}{Brian~T. Gill},
  \bibinfo{person}{Tadayoshi Kohno}, {and} \bibinfo{person}{William~H.
  Maisel}.} \bibinfo{year}{2010}\natexlab{}.
\newblock \showarticletitle{Patients, Pacemakers, and Implantable
  Defibrillators: Human Values and Security for Wireless Implantable Medical
  Devices}. \bibinfo{publisher}{Proceedings of the SIGCHI Conference on Human
  Factors in Computing Systems}.
\newblock


\bibitem[Digital(2019)]%
        {delloiteCA}
\bibfield{author}{\bibinfo{person}{Deloitte Digital}.}
  \bibinfo{year}{2019}\natexlab{}.
\newblock \bibinfo{title}{Conversational AI: The next wave of customer and
  employee experience}.
\newblock
\newblock


\bibitem[DigitalCatapult(2020)]%
        {digitalcatapult2020}
\bibfield{author}{\bibinfo{person}{DigitalCatapult}.}
  \bibinfo{year}{2020}\natexlab{}.
\newblock \bibinfo{title}{Lessons in Practical AI Ethics}.
\newblock
\newblock


\bibitem[D\'Ignazio and Klein(2020)]%
        {klein2020}
\bibfield{author}{\bibinfo{person}{C. D\'Ignazio} {and} \bibinfo{person}{L.
  Klein}.} \bibinfo{year}{2020}\natexlab{}.
\newblock \bibinfo{booktitle}{\emph{Data Feminism}}.
\newblock \bibinfo{publisher}{MIT Press}.
\newblock


\bibitem[Dindler et~al\mbox{.}(2022)]%
        {dindler2022}
\bibfield{author}{\bibinfo{person}{C. Dindler}, \bibinfo{person}{P. Krogh},
  \bibinfo{person}{K. Tikær}, {and} \bibinfo{person}{P. Nørregård}.}
  \bibinfo{year}{2022}\natexlab{}.
\newblock \showarticletitle{Engagements and articulations of ethics in design
  practice}.
\newblock \bibinfo{journal}{\emph{International Journal of Design}}
  \bibinfo{volume}{16}, \bibinfo{number}{2} (\bibinfo{year}{2022}),
  \bibinfo{pages}{47--56}.
\newblock
\urldef\tempurl%
\url{https://doi.org/10.57698/v16i2.04}
\showDOI{\tempurl}


\bibitem[Easton et~al\mbox{.}(2019)]%
        {easton2019}
\bibfield{author}{\bibinfo{person}{K. Easton}, \bibinfo{person}{S. Potter},
  \bibinfo{person}{R. Bec}, \bibinfo{person}{M. Bennion}, \bibinfo{person}{H.
  Christensen}, \bibinfo{person}{C. Grindell}, \bibinfo{person}{B. Mirheidari},
  \bibinfo{person}{S. Weich}, \bibinfo{person}{L. de Witte},
  \bibinfo{person}{D. Wolstenholme}, {and} \bibinfo{person}{M. Hawley}.}
  \bibinfo{year}{2019}\natexlab{}.
\newblock \showarticletitle{A Virtual Agent to Support Individuals Living With
  Physical and Mental Comorbidities: Co-Design and Acceptability Testing}.
\newblock \bibinfo{journal}{\emph{Journal of Medical Internet Research (JMIR)}}
  \bibinfo{volume}{21}, \bibinfo{number}{5} (\bibinfo{year}{2019}).
\newblock


\bibitem[Ehsan et~al\mbox{.}(2023)]%
        {ehsan2023}
\bibfield{author}{\bibinfo{person}{U. Ehsan}, \bibinfo{person}{K. Saha},
  \bibinfo{person}{M. Choudhury}, {and} \bibinfo{person}{M. Riedl}.}
  \bibinfo{year}{2023}\natexlab{}.
\newblock \showarticletitle{Charting the Sociotechnical Gap in Explainable AI:
  A Framework to Address the Gap in XAI}.
\newblock \bibinfo{journal}{\emph{Proceedings of the ACM on Human-Computer
  Interaction}} \bibinfo{number}{34} (\bibinfo{year}{2023}).
\newblock


\bibitem[Elshan et~al\mbox{.}(2022)]%
        {elshan2022}
\bibfield{author}{\bibinfo{person}{E. Elshan}, \bibinfo{person}{C. Engel},
  \bibinfo{person}{P. Ebel}, {and} \bibinfo{person}{D. Siemon}.}
  \bibinfo{year}{2022}\natexlab{}.
\newblock \showarticletitle{Assessing the Reusability of Design Principles in
  the Realm of Conversational Agents}. \bibinfo{publisher}{Proceedings of the
  International Conference on Design Science Research in Information Systems
  and Technology (DESRIST)}.
\newblock


\bibitem[Eubanks(2018)]%
        {eubanks2018}
\bibfield{author}{\bibinfo{person}{V. Eubanks}.}
  \bibinfo{year}{2018}\natexlab{}.
\newblock \bibinfo{booktitle}{\emph{Automating inequality: How high-tech tools
  profile, police, and punish the poor}}.
\newblock \bibinfo{publisher}{St. Martin's Press}.
\newblock


\bibitem[FAIR(2021)]%
        {fair2021}
\bibfield{author}{\bibinfo{person}{Go FAIR}.} \bibinfo{year}{2021}\natexlab{}.
\newblock \bibinfo{title}{Data Based Science: FAIR Becomes the New Normal}.
\newblock
\newblock
\urldef\tempurl%
\url{https://www.go-fair.org/2021/01/21/data-based-science-fair-becomes-the-new-normal/}
\showURL{%
\tempurl}


\bibitem[Flanagan et~al\mbox{.}(2008)]%
        {flanagan2008}
\bibfield{author}{\bibinfo{person}{M. Flanagan}, \bibinfo{person}{DC. Howe},
  {and} \bibinfo{person}{H. Nissenbaum}.} \bibinfo{year}{2008}\natexlab{}.
\newblock \showarticletitle{Embodying values in technology: Theory and
  practice}.
\newblock \bibinfo{journal}{\emph{Information technology and moral philosophy}}
  (\bibinfo{year}{2008}), \bibinfo{pages}{322--346}.
\newblock


\bibitem[Folstad et~al\mbox{.}(2021)]%
        {folstad2021}
\bibfield{author}{\bibinfo{person}{A. Folstad}, \bibinfo{person}{T. Araujo},
  \bibinfo{person}{E. Law}, \bibinfo{person}{P. Brandtzaeg},
  \bibinfo{person}{S. Papadopoulos}, \bibinfo{person}{L. Reis},
  \bibinfo{person}{M. Baez}, \bibinfo{person}{G. Laban}, \bibinfo{person}{P.
  McAllister}, \bibinfo{person}{C. Ischen}, \bibinfo{person}{R. Wald},
  \bibinfo{person}{F. Catania}, \bibinfo{person}{R. von Wolff},
  \bibinfo{person}{S. Hobert}, {and} \bibinfo{person}{E. Luger}.}
  \bibinfo{year}{2021}\natexlab{}.
\newblock \showarticletitle{Future directions for chatbot research: an
  interdisciplinary research agenda}.
\newblock \bibinfo{journal}{\emph{Computing}} (\bibinfo{year}{2021}),
  \bibinfo{pages}{2915--–2942}.
\newblock


\bibitem[F\o{}lstad et~al\mbox{.}(2018)]%
        {folstad2018}
\bibfield{author}{\bibinfo{person}{A. F\o{}lstad}, \bibinfo{person}{P.
  Brandtzaeg}, \bibinfo{person}{T. Feltwell}, \bibinfo{person}{E. Law},
  \bibinfo{person}{M. Tscheligi}, {and} \bibinfo{person}{E. Luger}.}
  \bibinfo{year}{2018}\natexlab{}.
\newblock \showarticletitle{SIG: Chatbots for Social Good}.
  \bibinfo{publisher}{Extended Abstracts of the CHI Conference on Human Factors
  in Computing Systems}.
\newblock


\bibitem[for AI~Safety(2023)]%
        {ai-warning}
\bibfield{author}{\bibinfo{person}{Centre for AI~Safety}.}
  \bibinfo{year}{2023}\natexlab{}.
\newblock \bibinfo{title}{Statement on AI Risk: AI experts and public figures
  express their concern about AI risk.}
\newblock
\newblock
\urldef\tempurl%
\url{https://www.safe.ai/statement-on-ai-risk}
\showURL{%
\tempurl}


\bibitem[Freier(2008)]%
        {freier2008}
\bibfield{author}{\bibinfo{person}{N. Freier}.}
  \bibinfo{year}{2008}\natexlab{}.
\newblock \showarticletitle{Children Attribute Moral Standing to a Personified
  Agent}. \bibinfo{publisher}{Proceedings of the SIGCHI Conference on Human
  Factors in Computing Systems}.
\newblock


\bibitem[Friedman et~al\mbox{.}(2021)]%
        {friedman2021}
\bibfield{author}{\bibinfo{person}{B. Friedman}, \bibinfo{person}{M. Harbers},
  \bibinfo{person}{D. Hendry}, \bibinfo{person}{J. van~den Hoven},
  \bibinfo{person}{C. Jonker}, {and} \bibinfo{person}{N. Logler}.}
  \bibinfo{year}{2021}\natexlab{}.
\newblock \showarticletitle{Eight grand challenges for value sensitive design
  from the 2016 Lorentz workshop}.
\newblock \bibinfo{journal}{\emph{Ethics and Information Technology}}
  \bibinfo{volume}{23} (\bibinfo{year}{2021}), \bibinfo{pages}{5--16}.
\newblock


\bibitem[Friedman and Hendry(2019)]%
        {friedmanbook}
\bibfield{author}{\bibinfo{person}{B. Friedman} {and} \bibinfo{person}{D.
  Hendry}.} \bibinfo{year}{2019}\natexlab{}.
\newblock \bibinfo{booktitle}{\emph{Value Sensitive Design: Shaping Technology
  with Moral Imagination}}.
\newblock \bibinfo{publisher}{MIT Press}.
\newblock


\bibitem[Friedman et~al\mbox{.}(2006)]%
        {friedman2006}
\bibfield{author}{\bibinfo{person}{B. Friedman}, \bibinfo{person}{P. Kahn},
  \bibinfo{person}{J. Hagman}, \bibinfo{person}{R. Severson}, {and}
  \bibinfo{person}{B. Gill}.} \bibinfo{year}{2006}\natexlab{}.
\newblock \showarticletitle{The Watcher and the Watched: Social Judgments About
  Privacy in a Public Place}.
\newblock \bibinfo{journal}{\emph{Human-Computer Interaction}}
  \bibinfo{volume}{21} (\bibinfo{year}{2006}), \bibinfo{pages}{235--272}.
\newblock


\bibitem[Fuchsberger et~al\mbox{.}(2012)]%
        {fuchsberger2012}
\bibfield{author}{\bibinfo{person}{V. Fuchsberger}, \bibinfo{person}{C. Moser},
  {and} \bibinfo{person}{M. Tscheligi}.} \bibinfo{year}{2012}\natexlab{}.
\newblock \showarticletitle{Values in Action (ViA): Combining Usability, User
  Experience and User Acceptance}. \bibinfo{publisher}{Proceedings of the
  SIGCHI Conference on Human Factors in Computing Systems (CHI)}.
\newblock


\bibitem[Garg and Sengupta(2020)]%
        {garg2020}
\bibfield{author}{\bibinfo{person}{R. Garg} {and} \bibinfo{person}{S.
  Sengupta}.} \bibinfo{year}{2020}\natexlab{}.
\newblock \showarticletitle{Conversational Technologies for In-home Learning:
  Using Co-Design to Understand Children’s and Parents’ Perspectives}.
  \bibinfo{publisher}{Proceedings of the CHI Conference on Human Factors in
  Computing Systems}.
\newblock


\bibitem[Görnemann and Spiekermann(2022)]%
        {gornemann2022}
\bibfield{author}{\bibinfo{person}{E. Görnemann} {and} \bibinfo{person}{S.
  Spiekermann}.} \bibinfo{year}{2022}\natexlab{}.
\newblock \showarticletitle{Emotional responses to human values in technology:
  The case of conversational agents}.
\newblock \bibinfo{journal}{\emph{Human–Computer Interaction}}
  (\bibinfo{year}{2022}), \bibinfo{pages}{1--28}.
\newblock


\bibitem[Hagendorff(2020)]%
        {hagendorff2020}
\bibfield{author}{\bibinfo{person}{T. Hagendorff}.}
  \bibinfo{year}{2020}\natexlab{}.
\newblock \showarticletitle{The ethics of AI ethics: an evaluation of
  guidelines}.
\newblock \bibinfo{journal}{\emph{Minds and Machines}}  \bibinfo{volume}{30}
  (\bibinfo{year}{2020}), \bibinfo{pages}{99--120}.
\newblock


\bibitem[Hallensleben and Hustedt(2020)]%
        {hallensleben2020}
\bibfield{author}{\bibinfo{person}{S. Hallensleben} {and} \bibinfo{person}{C.
  Hustedt}.} \bibinfo{year}{2020}\natexlab{}.
\newblock \bibinfo{title}{From Principles to Practice; An interdisciplinary
  framework to operationalise AI ethics}.
\newblock
\newblock


\bibitem[Harbers and Overdiek(2022)]%
        {harbers2022}
\bibfield{author}{\bibinfo{person}{M. Harbers} {and} \bibinfo{person}{A.
  Overdiek}.} \bibinfo{year}{2022}\natexlab{}.
\newblock \showarticletitle{Towards a living lab for responsible applied AI}.
  In \bibinfo{booktitle}{\emph{Proceedings of the Design Research Society
  Conference (DRS)}}.
\newblock


\bibitem[Hayes(2020)]%
        {Hayes2020}
\bibfield{author}{\bibinfo{person}{G. Hayes}.} \bibinfo{year}{2020}\natexlab{}.
\newblock \showarticletitle{Inclusive and Engaged HCI}.
\newblock \bibinfo{journal}{\emph{Interactions}} \bibinfo{volume}{27},
  \bibinfo{number}{2} (\bibinfo{year}{2020}), \bibinfo{pages}{26--31}.
\newblock
\urldef\tempurl%
\url{https://doi.org/10.1145/3378561}
\showDOI{\tempurl}


\bibitem[Hedman and Gimpel(2010)]%
        {hedman2010}
\bibfield{author}{\bibinfo{person}{J. Hedman} {and} \bibinfo{person}{G.
  Gimpel}.} \bibinfo{year}{2010}\natexlab{}.
\newblock \showarticletitle{The adoption of hyped technologies: a qualitative
  study}.
\newblock \bibinfo{journal}{\emph{Information and Technology Management}}
  \bibinfo{volume}{11}, \bibinfo{number}{4} (\bibinfo{year}{2010}),
  \bibinfo{pages}{161--175}.
\newblock


\bibitem[Heo and Lee(2023)]%
        {heo2023}
\bibfield{author}{\bibinfo{person}{J. Heo} {and} \bibinfo{person}{U. Lee}.}
  \bibinfo{year}{2023}\natexlab{}.
\newblock \showarticletitle{Form to Flow: Exploring Challenges and Roles of
  Conversational UX Designers in Real-world, Multi-channel Service
  Environments}. \bibinfo{publisher}{Proceedings of the ACM on Human-Computer
  Interaction}.
\newblock


\bibitem[Hockey(2007)]%
        {hockey2007}
\bibfield{author}{\bibinfo{person}{J. Hockey}.}
  \bibinfo{year}{2007}\natexlab{}.
\newblock \showarticletitle{United Kingdom art and design practice PhDs:
  Evidence from students and their supervisors}.
\newblock \bibinfo{journal}{\emph{Studies in Art Education}}
  \bibinfo{volume}{48}, \bibinfo{number}{2} (\bibinfo{year}{2007}),
  \bibinfo{pages}{155--170}.
\newblock


\bibitem[IEEE 7000-2021(2021)]%
        {ieee7000}
IEEE 7000-2021 \bibinfo{year}{2021}\natexlab{}.
\newblock \bibinfo{booktitle}{\emph{IEEE Standard Model Process for Addressing
  Ethical Concerns during System Design}}.
\newblock \bibinfo{type}{Standard}. \bibinfo{institution}{The Institute of
  Electrical and Electronics Engineers}.
\newblock


\bibitem[Jakesch et~al\mbox{.}(2022)]%
        {jakesch2022}
\bibfield{author}{\bibinfo{person}{M. Jakesch}, \bibinfo{person}{Z. Buçinca},
  \bibinfo{person}{S. Amershi}, {and} \bibinfo{person}{A. Olteanu}.}
  \bibinfo{year}{2022}\natexlab{}.
\newblock \showarticletitle{How Different Groups Prioritize Ethical Values for
  Responsible AI}. \bibinfo{publisher}{In Proceedings of the ACM Conference on
  Fairness, Accountability, and Transparency (FAccT)}.
\newblock


\bibitem[Jobin et~al\mbox{.}(2019)]%
        {jobin2019}
\bibfield{author}{\bibinfo{person}{A. Jobin}, \bibinfo{person}{M. Ienca}, {and}
  \bibinfo{person}{E. Vayena}.} \bibinfo{year}{2019}\natexlab{}.
\newblock \showarticletitle{The global landscape of AI ethics guidelines}.
\newblock \bibinfo{journal}{\emph{Natural Machine Intelligence}}
  \bibinfo{volume}{1} (\bibinfo{year}{2019}), \bibinfo{pages}{389–399}.
\newblock


\bibitem[Jonas(2018)]%
        {jonas2018}
\bibfield{author}{\bibinfo{person}{W. Jonas}.} \bibinfo{year}{2018}\natexlab{}.
\newblock \bibinfo{booktitle}{\emph{A Cybernetic Model of Design Research:
  Towards a trans-domain of knowing}}.
\newblock \bibinfo{publisher}{Routledge}.
\newblock


\bibitem[Koesten et~al\mbox{.}(2019)]%
        {koesten2019}
\bibfield{author}{\bibinfo{person}{L. Koesten}, \bibinfo{person}{E. Kacprzak},
  \bibinfo{person}{J. Tennison}, {and} \bibinfo{person}{E. Simperl}.}
  \bibinfo{year}{2019}\natexlab{}.
\newblock \showarticletitle{Collaborative Practices with Structured Data: Do
  Tools Support What Users Need?} \bibinfo{publisher}{Proceedings of the 2019
  CHI Conference on Human Factors in Computing Systems}.
\newblock


\bibitem[Koster et~al\mbox{.}(2022)]%
        {koster2022}
\bibfield{author}{\bibinfo{person}{R. Koster}, \bibinfo{person}{J. Balaguer},
  {and} \bibinfo{person}{A. Tacchetti}.} \bibinfo{year}{2022}\natexlab{}.
\newblock \showarticletitle{Human-centred mechanism design with Democratic AI}.
\newblock \bibinfo{journal}{\emph{Natural Human Behaviour}}
  \bibinfo{volume}{6} (\bibinfo{year}{2022}), \bibinfo{pages}{1398--1407}.
\newblock


\bibitem[Kot and Leszczyński(2020)]%
        {kot2020}
\bibfield{author}{\bibinfo{person}{M.T. Kot} {and} \bibinfo{person}{G.
  Leszczyński}.} \bibinfo{year}{2020}\natexlab{}.
\newblock \showarticletitle{The concept of intelligent agent in business
  interactions: is virtual assistant an actor or a boundary object?}
\newblock \bibinfo{journal}{\emph{Journal of Business \& Industrial Marketing}}
  \bibinfo{volume}{35}, \bibinfo{number}{7} (\bibinfo{year}{2020}),
  \bibinfo{pages}{1155--1164}.
\newblock


\bibitem[Kroes and Van~de Poel(2020)]%
        {kroes2020}
\bibfield{author}{\bibinfo{person}{P. Kroes} {and} \bibinfo{person}{I. Van~de
  Poel}.} \bibinfo{year}{2020}\natexlab{}.
\newblock \bibinfo{booktitle}{\emph{Design for Values and the Definition,
  Specification, and Operationalization of Values}}.
\newblock \bibinfo{publisher}{Springer Reference}.
\newblock


\bibitem[Lee et~al\mbox{.}(2017)]%
        {lee2017}
\bibfield{author}{\bibinfo{person}{S. Lee}, \bibinfo{person}{J. Lee}, {and}
  \bibinfo{person}{K. Lee}.} \bibinfo{year}{2017}\natexlab{}.
\newblock \showarticletitle{Designing Intelligent Assistant through User
  Participations}. \bibinfo{publisher}{Proceedings of the Conference on
  Designing Interactive Systems}.
\newblock


\bibitem[Lewandowski et~al\mbox{.}(2021)]%
        {lewandowski2021}
\bibfield{author}{\bibinfo{person}{T. Lewandowski}, \bibinfo{person}{J.
  Delling}, \bibinfo{person}{C. Grotherr}, {and} \bibinfo{person}{T.
  Böhmann}.} \bibinfo{year}{2021}\natexlab{}.
\newblock \showarticletitle{State-of-the-Art Analysis of Adopting AI-based
  Conversational Agents in Organizations: A Systematic Literature Review}.
  \bibinfo{publisher}{Proceedings of Pacific Asia Conference on Information
  Systems (PACIS)}.
\newblock


\bibitem[Lim and Kim(2022)]%
        {lim2022}
\bibfield{author}{\bibinfo{person}{Y. Lim} {and} \bibinfo{person}{B. Kim}.}
  \bibinfo{year}{2022}\natexlab{}.
\newblock \showarticletitle{Value-drive design approach to envision speculative
  futures}. In \bibinfo{booktitle}{\emph{Proceedings of the Design Research
  Society Conference (DRS)}}.
\newblock


\bibitem[Lopatovska et~al\mbox{.}(2022)]%
        {lopatovska2022}
\bibfield{author}{\bibinfo{person}{Irene Lopatovska}, \bibinfo{person}{Olivia
  Turpin}, \bibinfo{person}{Jessika Davis}, \bibinfo{person}{Ellen Connell},
  \bibinfo{person}{Chris Denney}, \bibinfo{person}{Hilda Fournier},
  \bibinfo{person}{Archana Ravi}, \bibinfo{person}{Ji~Hee Yoon}, {and}
  \bibinfo{person}{Eesha Parasnis}.} \bibinfo{year}{2022}\natexlab{}.
\newblock \showarticletitle{Capturing Teens’ Voice in Designing Supportive
  Agents}. \bibinfo{publisher}{Proceedings of the 4th Conference on
  Conversational User Interfaces}.
\newblock
\urldef\tempurl%
\url{https://doi.org/10.1145/3543829.3543838}
\showDOI{\tempurl}


\bibitem[Luria(2018)]%
        {luria2018}
\bibfield{author}{\bibinfo{person}{M. Luria}.} \bibinfo{year}{2018}\natexlab{}.
\newblock \showarticletitle{Designing Robot Personality Based on Fictional
  Sidekick Characters}. In \bibinfo{booktitle}{\emph{Proceedings of the
  ACM/IEEE International Conference on Human-Robot Interaction}}.
\newblock


\bibitem[Luria et~al\mbox{.}(2020)]%
        {luria2020}
\bibfield{author}{\bibinfo{person}{Michal Luria}, \bibinfo{person}{Judeth
  Oden~Choi}, \bibinfo{person}{Rachel~Gita Karp}, \bibinfo{person}{John
  Zimmerman}, {and} \bibinfo{person}{Jodi Forlizzi}.}
  \bibinfo{year}{2020}\natexlab{}.
\newblock \showarticletitle{Robotic Futures: Learning about Personally-Owned
  Agents through Performance}. \bibinfo{publisher}{Proceedings of the ACM
  Designing Interactive Systems Conference (DIS)}.
\newblock
\urldef\tempurl%
\url{https://doi.org/10.1145/3357236.3395488}
\showDOI{\tempurl}


\bibitem[Manders-Huits(2011)]%
        {manders2011}
\bibfield{author}{\bibinfo{person}{N. Manders-Huits}.}
  \bibinfo{year}{2011}\natexlab{}.
\newblock \showarticletitle{What Values in Design? The Challenge of
  Incorporating Moral Values into Design}.
\newblock \bibinfo{journal}{\emph{Science and Engineering Ethics}}
  \bibinfo{volume}{17}, \bibinfo{number}{2} (\bibinfo{year}{2011}),
  \bibinfo{pages}{271–--287}.
\newblock


\bibitem[Manders-Huits and Zimmer(2009)]%
        {manders2009}
\bibfield{author}{\bibinfo{person}{N. Manders-Huits} {and} \bibinfo{person}{M.
  Zimmer}.} \bibinfo{year}{2009}\natexlab{}.
\newblock \showarticletitle{Values and pragmatic action: The challenges of
  introducing ethical intelligence in technical design communities}.
\newblock \bibinfo{journal}{\emph{The International Review of Information
  Ethics}}  \bibinfo{volume}{10} (\bibinfo{year}{2009}),
  \bibinfo{pages}{37--44}.
\newblock


\bibitem[McNamara et~al\mbox{.}(2018)]%
        {mcnamara2018}
\bibfield{author}{\bibinfo{person}{A. McNamara}, \bibinfo{person}{J. Smith},
  {and} \bibinfo{person}{E. Murphy-Hill}.} \bibinfo{year}{2018}\natexlab{}.
\newblock \showarticletitle{Does ACM’s Code of Ethics Change Ethical Decision
  Making in Software Development?}. In \bibinfo{booktitle}{\emph{Proceedings of
  the 2018 26th ACM Joint Meeting on European Software Engineering Conference
  and Symposium on the Foundations of Software Engineering}} (Lake Buena Vista,
  FL, USA) \emph{(\bibinfo{series}{ESEC/FSE 2018})}.
  \bibinfo{publisher}{Association for Computing Machinery},
  \bibinfo{address}{New York, NY, USA}, \bibinfo{pages}{729–733}.
\newblock
\showISBNx{9781450355735}
\urldef\tempurl%
\url{https://doi.org/10.1145/3236024.3264833}
\showDOI{\tempurl}


\bibitem[Mepham et~al\mbox{.}(2006)]%
        {mepham2006}
\bibfield{author}{\bibinfo{person}{B. Mepham}, \bibinfo{person}{M. Kaiser},
  \bibinfo{person}{E. Thortensen}, \bibinfo{person}{S. Tomkins}, {and}
  \bibinfo{person}{K. Millar}.} \bibinfo{year}{2006}\natexlab{}.
\newblock \showarticletitle{Ethical Matrix Manual}.
\newblock \bibinfo{journal}{\emph{Agricultural and Forest Meteorology.}}
  (\bibinfo{year}{2006}).
\newblock


\bibitem[Miller et~al\mbox{.}(2007)]%
        {miller2007}
\bibfield{author}{\bibinfo{person}{J. Miller}, \bibinfo{person}{B. Friedman},
  \bibinfo{person}{G. Jancke}, {and} \bibinfo{person}{B. Gill}.}
  \bibinfo{year}{2007}\natexlab{}.
\newblock \showarticletitle{Value tensions in design: the value sensitive
  design, development, and appropriation of a corporation's groupware system}.
  \bibinfo{publisher}{Proceedings of the 2007 International ACM Conference on
  Supporting Group Work}.
\newblock


\bibitem[Moore(2018)]%
        {moore2018}
\bibfield{author}{\bibinfo{person}{R. Moore}.} \bibinfo{year}{2018}\natexlab{}.
\newblock \bibinfo{booktitle}{\emph{A natural conversation framework for
  conversational UX design}}.
\newblock \bibinfo{publisher}{Springer International Publishing},
  \bibinfo{pages}{181--204}.
\newblock


\bibitem[Moore and Arar(2019)]%
        {moore2019}
\bibfield{author}{\bibinfo{person}{R. Moore} {and} \bibinfo{person}{R. Arar}.}
  \bibinfo{year}{2019}\natexlab{}.
\newblock \bibinfo{booktitle}{\emph{Conversational UX Design: A Practitioner's
  Guide to the Natural Conversation Framework}}.
\newblock \bibinfo{publisher}{ACM Books}.
\newblock


\bibitem[Moore et~al\mbox{.}(2016)]%
        {moore2016}
\bibfield{author}{\bibinfo{person}{R. Moore}, \bibinfo{person}{R. Hosn}, {and}
  \bibinfo{person}{A. Arora}.} \bibinfo{year}{2016}\natexlab{}.
\newblock \showarticletitle{The machinery of natural conversation and the
  design of conversational machines}. \bibinfo{publisher}{American Sociological
  Association Annual Meeting}.
\newblock


\bibitem[Morley et~al\mbox{.}(2021)]%
        {morley2021}
\bibfield{author}{\bibinfo{person}{J. Morley}, \bibinfo{person}{L. Kinsey},
  \bibinfo{person}{A. Elhalal}, \bibinfo{person}{F. Garcia},
  \bibinfo{person}{M. Ziosi}, {and} \bibinfo{person}{L. Floridi}.}
  \bibinfo{year}{2021}\natexlab{}.
\newblock \showarticletitle{Operationalising AI ethics: barriers, enablers and
  next steps}.
\newblock \bibinfo{journal}{\emph{AI \& Society}} (\bibinfo{year}{2021}).
\newblock


\bibitem[Murad et~al\mbox{.}(2023)]%
        {murad2023}
\bibfield{author}{\bibinfo{person}{C. Murad}, \bibinfo{person}{H. Candello},
  {and} \bibinfo{person}{C. Munteanu}.} \bibinfo{year}{2023}\natexlab{}.
\newblock \showarticletitle{What’s The Talk on VUI Guidelines? A
  Meta-Analysis of Guidelines for Voice User Interface Design}.
  \bibinfo{publisher}{Proceedings of the ACM conference on Conversational User
  Interfaces (CUI)}.
\newblock


\bibitem[Ocnarescu et~al\mbox{.}(2011)]%
        {ocnarescu2011}
\bibfield{author}{\bibinfo{person}{I. Ocnarescu}, \bibinfo{person}{F. Pain},
  \bibinfo{person}{C. Bouchard}, \bibinfo{person}{A. Aoussat}, {and}
  \bibinfo{person}{D. Sciamma}.} \bibinfo{year}{2011}\natexlab{}.
\newblock \showarticletitle{Improvement of the Industrial Design Process by the
  Creation and Usage of Intermediate Representations of Technology,
  "TechCards"}. \bibinfo{publisher}{Proceedings of the 2011 Conference on
  Designing Pleasurable Products and Interfaces}.
\newblock


\bibitem[Paavola and Miettinen(2019)]%
        {paavola2019}
\bibfield{author}{\bibinfo{person}{S. Paavola} {and} \bibinfo{person}{R.
  Miettinen}.} \bibinfo{year}{2019}\natexlab{}.
\newblock \showarticletitle{Dynamics of Design Collaboration: BIM Models as
  Intermediary Digital Objects}.
\newblock \bibinfo{journal}{\emph{Computer Supported Cooperative Work (CSCW)}}
  \bibinfo{volume}{28} (\bibinfo{year}{2019}), \bibinfo{pages}{1--23}.
\newblock


\bibitem[Palmer and Schwan(2023)]%
        {palmer2023}
\bibfield{author}{\bibinfo{person}{A. Palmer} {and} \bibinfo{person}{D.
  Schwan}.} \bibinfo{year}{2023}\natexlab{}.
\newblock \showarticletitle{More Process, Less Principles: The Ethics of
  Deploying AI and Robotics in Medicine}.
\newblock \bibinfo{journal}{\emph{Cambridge Quarterly of Healthcare Ethics}}
  (\bibinfo{year}{2023}), \bibinfo{pages}{1--14}.
\newblock


\bibitem[Park et~al\mbox{.}(2022)]%
        {park2022}
\bibfield{author}{\bibinfo{person}{K. Park, J.and~Karahalios},
  \bibinfo{person}{N. Salehi}, {and} \bibinfo{person}{M. Eslami}.}
  \bibinfo{year}{2022}\natexlab{}.
\newblock \showarticletitle{Power Dynamics and Value Conflicts in Designing and
  Maintaining Socio-Technical Algorithmic Processes}.
\newblock   \bibinfo{volume}{6} (\bibinfo{year}{2022}).
\newblock


\bibitem[Pennington(2010)]%
        {pennington2010}
\bibfield{author}{\bibinfo{person}{D. Pennington}.}
  \bibinfo{year}{2010}\natexlab{}.
\newblock \showarticletitle{The Dynamics of Material Artifacts in Collaborative
  Research Teams}.
\newblock \bibinfo{journal}{\emph{Computer Supported Cooperative Work (CSCW)}}
  \bibinfo{volume}{19} (\bibinfo{year}{2010}), \bibinfo{pages}{175--199}.
\newblock


\bibitem[Piorkowski et~al\mbox{.}(2021)]%
        {piorkowski2021}
\bibfield{author}{\bibinfo{person}{D. Piorkowski}, \bibinfo{person}{S. Park},
  \bibinfo{person}{A. Wang}, \bibinfo{person}{D. Wang}, \bibinfo{person}{M.
  Muller}, {and} \bibinfo{person}{F. Portnoy}.}
  \bibinfo{year}{2021}\natexlab{}.
\newblock \showarticletitle{How AI Developers Overcome Communication Challenges
  in a Multidisciplinary Team: A Case Study}.
\newblock  (\bibinfo{year}{2021}).
\newblock


\bibitem[Pommeranz et~al\mbox{.}(2012)]%
        {pommeranz2012}
\bibfield{author}{\bibinfo{person}{A. Pommeranz}, \bibinfo{person}{C.
  Detweiler}, \bibinfo{person}{P. Wiggers}, {and} \bibinfo{person}{C. Jonker}.}
  \bibinfo{year}{2012}\natexlab{}.
\newblock \showarticletitle{Elicitation of situated values: need for tools to
  help stakeholders and designers to reflect and communicate}.
\newblock \bibinfo{journal}{\emph{Ethics and Information Technology}}
  \bibinfo{volume}{14} (\bibinfo{year}{2012}), \bibinfo{pages}{285--303}.
\newblock


\bibitem[Rahwan(2018)]%
        {rahwan2018}
\bibfield{author}{\bibinfo{person}{I. Rahwan}.}
  \bibinfo{year}{2018}\natexlab{}.
\newblock \showarticletitle{Society-in-the-loop: programming the algorithmic
  social contract}.
\newblock \bibinfo{journal}{\emph{Ethics and Information Technology}}
  \bibinfo{volume}{20}, \bibinfo{number}{1} (\bibinfo{year}{2018}),
  \bibinfo{pages}{5--14}.
\newblock


\bibitem[Redaelli and Carassa(2018)]%
        {redaelli2018}
\bibfield{author}{\bibinfo{person}{I. Redaelli} {and} \bibinfo{person}{A.
  Carassa}.} \bibinfo{year}{2018}\natexlab{}.
\newblock \showarticletitle{New Perspectives on Plans: Studying Planning as an
  Instance of Instructed Action}.
\newblock \bibinfo{journal}{\emph{Computer Supported Cooperative Work (CSCW)}}
  \bibinfo{volume}{27} (\bibinfo{year}{2018}), \bibinfo{pages}{107--148}.
\newblock


\bibitem[Russell(2019)]%
        {russell2019}
\bibfield{author}{\bibinfo{person}{S. Russell}.}
  \bibinfo{year}{2019}\natexlab{}.
\newblock \bibinfo{booktitle}{\emph{Human Compatible: AI and the Problem of
  Control}}.
\newblock \bibinfo{publisher}{Allen Lane}.
\newblock


\bibitem[Sadek et~al\mbox{.}(2023)]%
        {sadek2023}
\bibfield{author}{\bibinfo{person}{M. Sadek}, \bibinfo{person}{R.A. Calvo},
  {and} \bibinfo{person}{C. Mougenot}.} \bibinfo{year}{2023}\natexlab{}.
\newblock \showarticletitle{Trends, Challenges and Processes in Conversational
  Agent Design: Exploring Practitioners’ Views through Semi-Structured
  Interviews}. In \bibinfo{booktitle}{\emph{Proceedings of the 5th
  International Conference on Conversational User Interfaces}} (Eindhoven,
  Netherlands) \emph{(\bibinfo{series}{CUI '23})}.
  \bibinfo{publisher}{Association for Computing Machinery},
  \bibinfo{address}{New York, NY, USA}, Article \bibinfo{articleno}{13},
  \bibinfo{numpages}{10}~pages.
\newblock
\showISBNx{9798400700149}
\urldef\tempurl%
\url{https://doi.org/10.1145/3571884.3597143}
\showDOI{\tempurl}


\bibitem[Sanders and Stappers(2008)]%
        {strappers2008}
\bibfield{author}{\bibinfo{person}{E. Sanders} {and} \bibinfo{person}{P.
  Stappers}.} \bibinfo{year}{2008}\natexlab{}.
\newblock \showarticletitle{Co-creation and the new landscapes of design}.
\newblock \bibinfo{journal}{\emph{CoDesign}} \bibinfo{volume}{4},
  \bibinfo{number}{1} (\bibinfo{year}{2008}), \bibinfo{pages}{5--18}.
\newblock


\bibitem[Shilton(2013)]%
        {shilton2013}
\bibfield{author}{\bibinfo{person}{K. Shilton}.}
  \bibinfo{year}{2013}\natexlab{}.
\newblock \showarticletitle{Values Levers: Building Ethics into Design}.
\newblock \bibinfo{journal}{\emph{Science, Technology, \& Human Values}}
  \bibinfo{volume}{38}, \bibinfo{number}{3} (\bibinfo{year}{2013}),
  \bibinfo{pages}{374--397}.
\newblock


\bibitem[Showkat and Baumer(2022)]%
        {showkat2022}
\bibfield{author}{\bibinfo{person}{D. Showkat} {and} \bibinfo{person}{E.
  Baumer}.} \bibinfo{year}{2022}\natexlab{}.
\newblock \showarticletitle{“It’s Like the Value System in the Loop”:
  Domain Experts’ Values Expectations for NLP Automation}.
  \bibinfo{publisher}{Proceedings of the Designing Interactive Systems (DIS)
  Conference}.
\newblock


\bibitem[Sleeswijk(2018)]%
        {sleeswijk2018}
\bibfield{author}{\bibinfo{person}{V. Sleeswijk}.}
  \bibinfo{year}{2018}\natexlab{}.
\newblock \showarticletitle{Structuring Roles in Research through Design
  Collaboration}. \bibinfo{publisher}{Proceedings of the Design Research
  Society, DRS 2018: Design as a catalyst for change}.
\newblock


\bibitem[Spatz(2018)]%
        {spatz2018}
\bibfield{author}{\bibinfo{person}{J. Spatz}.} \bibinfo{year}{2018}\natexlab{}.
\newblock \bibinfo{title}{Skeuomorphism in VUI design: Jennifer Spatz from
  Rhode Island School of Design}.
\newblock
\newblock
\urldef\tempurl%
\url{https://www.youtube.com/watch?v=VxIrn5AV2Eo}
\showURL{%
\tempurl}


\bibitem[Spiekermann(2021)]%
        {spiekermann2021}
\bibfield{author}{\bibinfo{person}{S. Spiekermann}.}
  \bibinfo{year}{2021}\natexlab{}.
\newblock \showarticletitle{From value-lists to value-based engineering with
  IEEE 7000™}. In \bibinfo{booktitle}{\emph{IEEE International Symposium on
  Technology and Society (ISTAS)}}.
\newblock


\bibitem[Steen(2013)]%
        {steen2013}
\bibfield{author}{\bibinfo{person}{M. Steen}.} \bibinfo{year}{2013}\natexlab{}.
\newblock \showarticletitle{Co-Design as a Process of Joint Inquiry and
  Imagination}.
\newblock \bibinfo{journal}{\emph{Design Issues}}  \bibinfo{volume}{29}
  (\bibinfo{year}{2013}), \bibinfo{pages}{16--28}.
\newblock


\bibitem[Stray et~al\mbox{.}(2020)]%
        {stray2020}
\bibfield{author}{\bibinfo{person}{J. Stray}, \bibinfo{person}{I. Vendrov},
  \bibinfo{person}{J. Nixon}, \bibinfo{person}{S. Adler}, {and}
  \bibinfo{person}{D. Hadfield-Menell}.} \bibinfo{year}{2020}\natexlab{}.
\newblock \showarticletitle{What are you optimizing for? Aligning Recommender
  Systems with Human Values}. \bibinfo{publisher}{Proceedings of the ICML 2020
  Participatory Approaches to Machine Learning workshop}.
\newblock


\bibitem[Strikwerda et~al\mbox{.}(2022)]%
        {strikwerda2022}
\bibfield{author}{\bibinfo{person}{L. Strikwerda}, \bibinfo{person}{M. van
  Steenbergen}, \bibinfo{person}{A. van Gorp}, \bibinfo{person}{C. Timmers},
  {and} \bibinfo{person}{J. van Grondelle}.} \bibinfo{year}{2022}\natexlab{}.
\newblock \showarticletitle{The value sensitive design of a preventive health
  check app}.
\newblock \bibinfo{journal}{\emph{Ethics and Information Technology}}
  \bibinfo{volume}{24}, \bibinfo{number}{38} (\bibinfo{year}{2022}).
\newblock


\bibitem[Umbrello(2018)]%
        {umbrello2018}
\bibfield{author}{\bibinfo{person}{S. Umbrello}.}
  \bibinfo{year}{2018}\natexlab{}.
\newblock \showarticletitle{The moral psychology of value sensitive design: the
  methodological issues of moral intuitions for responsible innovation}.
\newblock \bibinfo{journal}{\emph{Journal of Responsible Innovation}}
  \bibinfo{volume}{5}, \bibinfo{number}{2} (\bibinfo{year}{2018}),
  \bibinfo{pages}{186--200}.
\newblock
\urldef\tempurl%
\url{https://doi.org/10.1080/23299460.2018.1457401}
\showDOI{\tempurl}


\bibitem[Umbrello et~al\mbox{.}(2021)]%
        {umbrello2021a}
\bibfield{author}{\bibinfo{person}{S. Umbrello}, \bibinfo{person}{M. Capasso},
  \bibinfo{person}{M. Balistreri}, \bibinfo{person}{A. Pirni}, {and}
  \bibinfo{person}{F. Merenda}.} \bibinfo{year}{2021}\natexlab{}.
\newblock \showarticletitle{Value Sensitive Design to Achieve the UN SDGs with
  AI: A Case of Elderly Care Robots}.
\newblock \bibinfo{journal}{\emph{Mind and Machines}}  \bibinfo{volume}{31}
  (\bibinfo{year}{2021}), \bibinfo{pages}{395--419}.
\newblock


\bibitem[Umbrello and Van~de Poel(2021)]%
        {umbrello2021}
\bibfield{author}{\bibinfo{person}{S. Umbrello} {and} \bibinfo{person}{I.
  Van~de Poel}.} \bibinfo{year}{2021}\natexlab{}.
\newblock \showarticletitle{Mapping value sensitive design onto AI for social
  good principles}.
\newblock \bibinfo{journal}{\emph{AI and Ethics}} \bibinfo{volume}{1},
  \bibinfo{number}{3} (\bibinfo{year}{2021}), \bibinfo{pages}{283--296}.
\newblock


\bibitem[van~de Poel(2013)]%
        {poel2013}
\bibfield{author}{\bibinfo{person}{I. van~de Poel}.}
  \bibinfo{year}{2013}\natexlab{}.
\newblock \bibinfo{booktitle}{\emph{Translating Values into Design
  Requirements}}.
\newblock \bibinfo{publisher}{Springer}.
\newblock


\bibitem[van~de Poel(2020)]%
        {poel2020}
\bibfield{author}{\bibinfo{person}{I. van~de Poel}.}
  \bibinfo{year}{2020}\natexlab{}.
\newblock \showarticletitle{Embedding Values in Artificial Intelligence (AI)
  Systems}.
\newblock \bibinfo{journal}{\emph{Mind and Machines}}  \bibinfo{volume}{30}
  (\bibinfo{year}{2020}), \bibinfo{pages}{385--409}.
\newblock


\bibitem[Van~de Poel and Kroes(2014)]%
        {poel2014}
\bibfield{author}{\bibinfo{person}{I. Van~de Poel} {and} \bibinfo{person}{P.
  Kroes}.} \bibinfo{year}{2014}\natexlab{}.
\newblock \bibinfo{booktitle}{\emph{Can Technology Embody Values?}}
\newblock \bibinfo{publisher}{Springer}, \bibinfo{pages}{103--124}.
\newblock


\bibitem[Varanasi and Goyal(2023)]%
        {varanasi2023}
\bibfield{author}{\bibinfo{person}{Rama~Adithya Varanasi} {and}
  \bibinfo{person}{Nitesh Goyal}.} \bibinfo{year}{2023}\natexlab{}.
\newblock \showarticletitle{“It is Currently Hodgepodge”: Examining AI/ML
  Practitioners’ Challenges during Co-Production of Responsible AI Values}.
  \bibinfo{publisher}{Proceedings of the CHI Conference on Human Factors in
  Computing Systems}.
\newblock
\urldef\tempurl%
\url{https://doi.org/10.1145/3544548.3580903}
\showDOI{\tempurl}


\bibitem[Vera and Muller(2019)]%
        {vera2019}
\bibfield{author}{\bibinfo{person}{Q. Vera} {and} \bibinfo{person}{M. Muller}.}
  \bibinfo{year}{2019}\natexlab{}.
\newblock \showarticletitle{Enabling Value Sensitive {AI} Systems through
  Participatory Design Fictions}.
\newblock \bibinfo{journal}{\emph{CoRR}} (\bibinfo{year}{2019}).
\newblock


\bibitem[Verdiesen and Dignum(2022)]%
        {verdiesen2022}
\bibfield{author}{\bibinfo{person}{I. Verdiesen} {and} \bibinfo{person}{V.
  Dignum}.} \bibinfo{year}{2022}\natexlab{}.
\newblock \showarticletitle{Value elicitation on a scenario of autonomous
  weapon system deployment: a qualitative study based on the value deliberation
  process}.
\newblock \bibinfo{journal}{\emph{AI and Ethics}} (\bibinfo{year}{2022}).
\newblock


\bibitem[Vermaas et~al\mbox{.}(2014)]%
        {vermaas2014}
\bibfield{author}{\bibinfo{person}{P. Vermaas}, \bibinfo{person}{P. Hekkert},
  \bibinfo{person}{N. Mander-Huits}, {and} \bibinfo{person}{N. Tromp}.}
  \bibinfo{year}{2014}\natexlab{}.
\newblock \bibinfo{booktitle}{\emph{Design Methods in Design for Values}}.
\newblock \bibinfo{publisher}{Springer}.
\newblock


\bibitem[Vines et~al\mbox{.}(2013)]%
        {vines2013}
\bibfield{author}{\bibinfo{person}{J. Vines}, \bibinfo{person}{R. Clarke},
  \bibinfo{person}{P. Wright}, \bibinfo{person}{J. McCarthy}, {and}
  \bibinfo{person}{P. Olivier}.} \bibinfo{year}{2013}\natexlab{}.
\newblock \showarticletitle{Configuring Participation: On How We Involve People
  in Design}. \bibinfo{publisher}{Proceedings of the SIGCHI Conference on Human
  Factors in Computing Systems}.
\newblock


\bibitem[Wachter and Mittelstadt(2019)]%
        {wachter2019}
\bibfield{author}{\bibinfo{person}{S. Wachter} {and} \bibinfo{person}{B.
  Mittelstadt}.} \bibinfo{year}{2019}\natexlab{}.
\newblock \showarticletitle{A Right to Reasonable Inferences: Re-Thinking Data
  Protection Law in the Age of Big Data and AI}.
\newblock \bibinfo{journal}{\emph{Columbia Business Law Review}}
  \bibinfo{number}{2} (\bibinfo{year}{2019}).
\newblock


\bibitem[Wahde and Virgolin(2022)]%
        {wahde2022}
\bibfield{author}{\bibinfo{person}{M. Wahde} {and} \bibinfo{person}{M.
  Virgolin}.} \bibinfo{year}{2022}\natexlab{}.
\newblock \bibinfo{booktitle}{\emph{Conversational Agents: Theory and
  Applications}}.
\newblock \bibinfo{publisher}{World Scientific Publishing Company}.
\newblock


\bibitem[Wambsganss et~al\mbox{.}(2021)]%
        {wambsganss2021}
\bibfield{author}{\bibinfo{person}{T. Wambsganss}, \bibinfo{person}{A.
  H{\"o}ch}, \bibinfo{person}{N. Zierau}, {and} \bibinfo{person}{M.
  S{\"o}llner}.} \bibinfo{year}{2021}\natexlab{}.
\newblock \showarticletitle{Ethical Design of Conversational Agents: Towards
  Principles for a Value-Sensitive Design}.
\newblock \bibinfo{journal}{\emph{Lecture Notes in Information Systems and
  Organisation}} (\bibinfo{year}{2021}).
\newblock


\bibitem[Weber et~al\mbox{.}(2021)]%
        {weber2021}
\bibfield{author}{\bibinfo{person}{P. Weber}, \bibinfo{person}{K. Krings},
  \bibinfo{person}{J. Nie\ss{}ner}, \bibinfo{person}{S. Brodesser}, {and}
  \bibinfo{person}{T. Ludwig}.} \bibinfo{year}{2021}\natexlab{}.
\newblock \showarticletitle{FoodChattAR: Exploring the Design Space of Edible
  Virtual Agents for Human-Food Interaction}. \bibinfo{publisher}{Proceedings
  of the ACM Designing Interactive Systems Conference (DIS)}.
\newblock
\urldef\tempurl%
\url{https://doi.org/10.1145/3461778.3461998}
\showDOI{\tempurl}


\bibitem[West et~al\mbox{.}(2019)]%
        {ainow2019}
\bibfield{author}{\bibinfo{person}{S. West}, \bibinfo{person}{M. Whittaker},
  {and} \bibinfo{person}{K. Crawford}.} \bibinfo{year}{2019}\natexlab{}.
\newblock \bibinfo{title}{Discriminating Systems: Gender, Race, and Power}.
\newblock
\newblock


\bibitem[Whittlestone et~al\mbox{.}(2019a)]%
        {whittlestone2019b}
\bibfield{author}{\bibinfo{person}{J. Whittlestone}, \bibinfo{person}{R.
  Nyrup}, \bibinfo{person}{A. Alexandrova}, {and} \bibinfo{person}{S. Cave}.}
  \bibinfo{year}{2019}\natexlab{a}.
\newblock \showarticletitle{The Role and Limits of Principles in AI Ethics:
  Towards a Focus on Tensions}. \bibinfo{publisher}{Proceedings of the AAAI/ACM
  Conference on AI, Ethics, and Society}.
\newblock


\bibitem[Whittlestone et~al\mbox{.}(2019b)]%
        {whittlestone2019c}
\bibfield{author}{\bibinfo{person}{J. Whittlestone}, \bibinfo{person}{R.
  Nyrup}, \bibinfo{person}{A. Alexandrova}, {and} \bibinfo{person}{S. Cave}.}
  \bibinfo{year}{2019}\natexlab{b}.
\newblock \showarticletitle{The Role and Limits of Principles in AI Ethics:
  Towards a Focus on Tensions}. In \bibinfo{booktitle}{\emph{Proceedings of the
  AAAI/ACM Conference on AI, Ethics, and Society}}.
\newblock


\bibitem[Winecoff and Watkins(2022)]%
        {winecoff2022}
\bibfield{author}{\bibinfo{person}{A. Winecoff} {and} \bibinfo{person}{E.
  Watkins}.} \bibinfo{year}{2022}\natexlab{}.
\newblock \showarticletitle{Artificial Concepts of Artificial Intelligence:
  Institutional Compliance and Resistance in AI Startups}.
  \bibinfo{publisher}{Proceedings of the AAAI/ACM Conference on AI, Ethics, and
  Society}.
\newblock
\urldef\tempurl%
\url{https://doi.org/10.1145/3514094.3534138}
\showDOI{\tempurl}


\bibitem[Woelfer and Hendry(2009)]%
        {woelfer2009}
\bibfield{author}{\bibinfo{person}{J. Woelfer} {and} \bibinfo{person}{D.
  Hendry}.} \bibinfo{year}{2009}\natexlab{}.
\newblock \showarticletitle{Stabilizing homeless young people with information
  and place}.
\newblock \bibinfo{journal}{\emph{Journal of the American Society for
  Information Science and Technology}} \bibinfo{volume}{60},
  \bibinfo{number}{11} (\bibinfo{year}{2009}), \bibinfo{pages}{2300--2312}.
\newblock


\bibitem[Wong et~al\mbox{.}(2022)]%
        {wong2022}
\bibfield{author}{\bibinfo{person}{R. Wong}, \bibinfo{person}{M. Madaio}, {and}
  \bibinfo{person}{N. Merrill}.} \bibinfo{year}{2022}\natexlab{}.
\newblock \showarticletitle{Seeing Like a Toolkit: How Toolkits Envision the
  Work of AI Ethics}.
\newblock \bibinfo{journal}{\emph{Computing Research Repository (CoRR)}}
  (\bibinfo{year}{2022}).
\newblock
\urldef\tempurl%
\url{https://arxiv.org/abs/2202.08792}
\showURL{%
\tempurl}


\bibitem[Yang et~al\mbox{.}(2018)]%
        {yang2018}
\bibfield{author}{\bibinfo{person}{Q. Yang}, \bibinfo{person}{A. Scuito},
  \bibinfo{person}{J. Zimmerman}, \bibinfo{person}{J. Forlizzi}, {and}
  \bibinfo{person}{A. Steinfeld}.} \bibinfo{year}{2018}\natexlab{}.
\newblock \showarticletitle{Investigating How Experienced UX Designers
  Effectively Work with Machine Learning}. \bibinfo{publisher}{Proceedings of
  the CHI Conference on Designing Interactive Systems (DIS)}.
\newblock


\bibitem[Yildirim et~al\mbox{.}(2023)]%
        {yildirim2023}
\bibfield{author}{\bibinfo{person}{N. Yildirim}, \bibinfo{person}{M.
  Pushkarna}, \bibinfo{person}{N. Goyal}, \bibinfo{person}{M. Wattenberg},
  {and} \bibinfo{person}{F. Viégas}.} \bibinfo{year}{2023}\natexlab{}.
\newblock \showarticletitle{Investigating How Practitioners Use Human-AI
  Guidelines: A Case Study on the People + AI Guidebook}.
  \bibinfo{publisher}{In Proceedings of the CHI Conference on Human Factors in
  Computing Systems,}.
\newblock


\bibitem[Yoo et~al\mbox{.}(2013)]%
        {yoo2013}
\bibfield{author}{\bibinfo{person}{D. Yoo}, \bibinfo{person}{A. Huldtgren},
  \bibinfo{person}{J. Woelfer}, \bibinfo{person}{D. Hendry}, {and}
  \bibinfo{person}{B. Friedman}.} \bibinfo{year}{2013}\natexlab{}.
\newblock \showarticletitle{A value sensitive action-reflection model: evolving
  a co-design space with stakeholder and designer prompts}.
  \bibinfo{publisher}{Proceedings of the SIGCHI Conference on Human Factors in
  Computing Systems (CHI)}.
\newblock


\bibitem[Yurrita et~al\mbox{.}(2022)]%
        {yurrita2022}
\bibfield{author}{\bibinfo{person}{M. Yurrita}, \bibinfo{person}{D.
  Murray-Rust}, \bibinfo{person}{A. Balayn}, {and} \bibinfo{person}{A.
  Bozzon}.} \bibinfo{year}{2022}\natexlab{}.
\newblock \showarticletitle{Towards a multi-stakeholder value-based assessment
  framework for algorithmic systems}. \bibinfo{publisher}{Proceedings of the
  ACM Conference on Fairness, Accountability, and Transparency (FAccT)}.
\newblock


\bibitem[Zhao et~al\mbox{.}(2023)]%
        {zhao2023}
\bibfield{author}{\bibinfo{person}{S. Zhao}, \bibinfo{person}{F. Tan}, {and}
  \bibinfo{person}{K. Fennedy}.} \bibinfo{year}{2023}\natexlab{}.
\newblock \showarticletitle{Heads-Up Computing Moving Beyond the
  Device-Centered Paradigm}.
\newblock \bibinfo{journal}{\emph{Commun. ACM}} \bibinfo{volume}{66},
  \bibinfo{number}{9} (\bibinfo{date}{aug} \bibinfo{year}{2023}),
  \bibinfo{pages}{56–63}.
\newblock
\showISSN{0001-0782}
\urldef\tempurl%
\url{https://doi.org/10.1145/3571722}
\showDOI{\tempurl}


\end{thebibliography}

\appendix

\end{document}